\documentclass[11pt]{article}
\usepackage[utf8]{inputenc}

\usepackage{graphicx}
\usepackage{scrextend}
\usepackage{amsthm}
\usepackage{comment}
\usepackage{graphicx,marvosym}
\usepackage{float}
\usepackage{hyperref}
\usepackage[english]{babel}
\usepackage{amsthm}

\usepackage[colorinlistoftodos,textsize=tiny,textwidth=2cm,color=red!25!white,obeyFinal]{todonotes}

\usepackage{geometry}
\usepackage{graphicx,adjustbox}
\usepackage{amsmath}
\usepackage{amssymb}
\usepackage[all]{xy}
\usepackage{tikz}
\usepackage[utf8]{inputenc}
\usepackage{csquotes}
\usepackage{cite}
\usepackage{lipsum} 
\usepackage[titletoc,toc,title]{appendix}

\usepackage{arydshln}
\usepackage{xcolor}
\usepackage{graphicx}
\usepackage{subcaption, caption}
\usepackage{multirow}
\usepackage{tikz}  
\usetikzlibrary{arrows,matrix,positioning}
\usetikzlibrary{matrix,fit,calc,positioning,arrows,chains,scopes,decorations.pathreplacing}
\newcommand\mvline[3][]{%
  \pgfmathtruncatemacro\hc{#3-1}
  \draw[#1]({$(#2-1-#3)!.5!(#2-1-\hc)$} |- #2.north) -- ({$(#2-1-#3)!.5!(#2-1-\hc)$} |- #2.south);
}

\newtheorem{theorem}{Theorem}[section]

\newtheorem{lemma}[theorem]{Lemma}
\newtheorem{proposition}[theorem]{Proposition}

\newtheorem{example}[theorem]{Example}

\newtheorem{claim}[theorem]{Claim}

\makeatletter
\renewcommand*\env@matrix[1][*\c@MaxMatrixCols c]{%
  \hskip -\arraycolsep
  \let\@ifnextchar\new@ifnextchar
  \array{#1}}
\makeatother

\begin{document}
\title {Improved Lower Bounds for Truthful Scheduling} 
\author{Shahar Dobzinski \and Ariel Shaulker\thanks{Weizmann Institute of Science.  Emails: {\ttfamily\{shahar.dobzinski, ariel.shaulker\}@weizmann.ac.il}. Work supported by BSF grant 2016192 and ISF grant 2185/19.}}

\maketitle

\begin{abstract}
The problem of scheduling unrelated machines by a truthful mechanism to minimize the makespan was introduced in the seminal ``Algorithmic Mechanism Design'' paper by Nisan and Ronen. Nisan and Ronen showed that there is a truthful mechanism that provides an approximation ratio of $\min(m,n)$, where $n$ is the number of machines and $m$ is the number of jobs. They also proved that no truthful mechanism can provide an approximation ratio better than $2$. Since then, the lower bound was improved to $1 +\sqrt 2 \approx 2.41$ by Christodoulou, Kotsoupias, and Vidali, and then to $1+\phi\approx 2.618$ by Kotsoupias and Vidali. Very recently, the lower bound was improved to $2.755$ by Giannakopoulos, Hammerl, and Pocas. In this paper we further improve the bound to $3-\delta$, for every constant $\delta>0$.

Note that a gap between the upper bound and the lower bounds exists even when the number of machines and jobs is very small. In particular, the known $1+\sqrt{2}$ lower bound requires at least $3$ machines and $5$ jobs. In contrast, we show a lower bound of $2.2055$ that uses only $3$ machines and $3$ jobs and a lower bound of $1+\sqrt 2$ that uses only $3$ machines and $4$ jobs. For the case of two machines and two jobs we show a lower bound of $2$. Similar bounds for two machines and two jobs were known before but only via complex proofs that characterized all truthful mechanisms that provide a finite approximation ratio in this setting, whereas our new proof uses a simple and direct approach.
\end{abstract}    


\section{Introduction}

We consider the problem of scheduling unrelated machines to minimize the makespan by a truthful mechanism. In this problem we have $n$ machines and $m$ jobs. Denote by $t^j_i$ the time it takes machine $i$ to process job $j$. The $t^j_i$'s are the private information of machine $i$. The goal is to minimize the makespan by a truthful mechanism, that is, find an allocation $(x_1,\ldots, x_n)$ of all jobs such that $\max_{i}\sum_{j\in x_i}t^j_i$ is minimized.

The problem was introduced by Nisan and Ronen in their seminal ``Algorithmic Mechanism Design'' paper \cite{NR99}. Nisan and Ronen showed that the VCG mechanism provides an approximation ratio of $\min (m,n)$. They also proved a lower bound of $2$  on the approximation ratio. Closing this gap is a major open question that has attracted much attention. 

The first improvement over these bounds was obtained by Christodoulou, Kotsoupias, and Vidali \cite{CKV07} who improved the bound to $1+\sqrt 2\approx 2.41$. Kotsoupias and Vidali \cite{KV} further improved the bound to $1+\phi\approx 2.618$. No improvement over this bound was obtained for more than a decade until very recently when Giannakopoulos, Hammerl, and Pocas \cite{GHP20} were able to improve the bound to $2.755$.

The gap between the lower and upper bounds is obviously still very large. Evidence that the ``correct'' answer is a lower bound of $n$ was provided by Ashlagi et al. \cite{ADL09} who showed that no truthful \emph{anonymous} mechanism can guarantee an approximation ratio better than $n$.

Generalizations of this problem were also studied. In particular, when the valuations are submodular (and not just additive), Christodoulou, Kotsoupias, and Kovacs obtain a lower bound of $\Omega(\sqrt n)$ \cite{CKK19}. Many papers also considered randomized and fractional versions of the problem \cite{NR99, MS18, YU09, LY08a, LU09, CDZ15, KV19, LS09, CKK10}.

\emph{The main result of this paper is an improvement of the lower bound to $3-\delta$, for every constant $\delta>0$.} We achieve this bound by carefully refining the constructions of \cite{KV, GHP20}. This result is provided in Section \ref{main_setion}.

We then go on by considering instances with small numbers of jobs and machines. Note that the VCG mechanism and slight modifications of it are the only mechanisms that we know that provide a finite approximation ratio. In fact, mechanisms that achieve non--trivial approximation guarantees are unknown even for very small instances. For example, the $1+\sqrt{2}$ lower bound of \cite{CKV07} requires three machines and five jobs. Currently, we do not even know whether there are mechanisms that provide an approximation ratio better than $3$ for instances with only three jobs and three machines.

In this paper we make progress in understanding the power of truthful mechanisms for small instances. We show a lower bound of $2.2055$ for instances with only $3$ machines and $3$ jobs (Section \ref{3*3}). In Section \ref{3*4} we also provide a lower bound of $1+\sqrt 2$ that uses only $3$ machines and $4$ jobs (this matches the lower bound of \cite{CKV07} that achieves the same bound with $3$ machines and $5$ jobs). We also consider instances with two machines and two jobs and show a lower bound of $2$ (Section \ref{2*2}). Similar bounds for two machines and two jobs were known before \cite{DS08, CKV08} but only via complex proofs that characterized all truthful mechanisms that provide a finite approximation ratio in this setting, whereas our new proof uses a simple and direct approach which is much more in line with all other lower bounds.

In Section~\ref{unified_view} we provide a new high--level view of the previous lower bounds  \cite{CKV07, KV, GHP20, NR99}. The new view allows us to give some intuition to the proof of the $3-\delta$ lower bound that is provided in Section~\ref{intuition_3}. In Section~\ref{warm-up} we use the new view and sketch the proof of a special case of the main theorem that yields a lower bound of $2.8019$. 

\subsection{A New High--Level View of Previous Lower Bounds} \label{unified_view}

We observe that the initial instances used in the proofs of the previous lower bounds \cite{CKV07, KV, GHP20, NR99} can be seen as a concatenation of at most three matrices -- $\mathcal{B}, \mathcal{C}$ and $\mathcal{D}$.  $\mathcal{D}$ is a matrix of dummy--jobs for all players (a matrix with zeroes on its diagonal and $\infty$ everywhere else, as depicted in Figure~\ref{BCD}), $\mathcal{C}$ is described next and $\mathcal{B}$ is the remainder of the instance. A $\mathcal{C}$ matrix is a matrix with the form depicted in Figure~\ref{BCD}. If we multiply all costs in a $\mathcal{C}$ matrix by the same factor or add players with cost of $\infty$ for all jobs then the resulting matrix is also considered a $\mathcal{C}$ matrix.

\begin{figure}[H]
    \centering
\begin{equation*}
        \mathcal{C}=
\begin{tikzpicture}[baseline=(m1-3-1.base)]
\matrix [matrix of math nodes,left delimiter=(,right delimiter=)] (m1)
{
\frac{1}{a^2}  &  \frac{1}{a^3}  &  \dots & \frac{1}{a^i} & \dots &\frac{1}{a^n}\\
\frac{1}{a}  &  \infty &\dots & \dots &\dots &\dots\\
\infty  &  \frac{1}{a^2} &\infty & \dots & \dots & \dots\\
\vdots &\vdots &\ddots &\vdots &\vdots &\vdots\\
\infty & \dots &\dots &\frac{1}{a^{i-1}} &\infty &\dots\\
\vdots &\vdots &\vdots &\vdots &\ddots &\vdots\\
\infty & \dots &\dots &\dots &\dots &\frac{1}{a^{n-1}}\\
};
\end{tikzpicture}
        \mathcal{D}=
\begin{tikzpicture}[baseline=(m1-3-1.base)]
\matrix [matrix of math nodes,left delimiter=(,right delimiter=)] (m1)
{
0       &  \infty  &  \dots & \dots & \dots &\dots &\dots\\
\infty  &    0     & \infty & \dots & \dots &\dots &\dots\\
\vdots  &  \vdots  & \ddots & \vdots&\vdots &\vdots&\vdots\\
\vdots  &  \vdots  & \vdots & \ddots&\vdots &\vdots&\vdots\\
\vdots  &  \vdots  & \vdots & \vdots&\ddots &\vdots&\vdots\\
\vdots  &  \vdots  & \vdots & \vdots&\vdots &\ddots&\vdots\\
\infty  &\dots     &\dots   &\dots  &\dots  &\dots &0\\
};
\end{tikzpicture}
\end{equation*}
    \caption{The matrices C and D}
    \label{BCD}
\end{figure}

Next, we describe the role of the $\mathcal{D}$ matrix and introduce a key claim regarding an instance with all three matrices $\mathcal{BCD}$ that will be used in the proofs of the previous lower bounds. The Matrix $\mathcal{D}$ is discussed in Section~\ref{D} and the claim is presented in Section~\ref{bcd_instance}. In Section~\ref{previous_proofs} we will see how to derive the previous bounds by applying our new point of view.

\subsubsection{The Matrix $\mathcal{D}$}\label{D}

Let $T$ be a matrix with $n \geq 2$ players and $\mathcal{A}$ the set of valid allocations \footnote{Observe that any mechanism with finite approximation ratio must allocate each player his dummy job. For this reason, we will assume that all the allocations we consider satisfy it.}. 
Consider an allocation $A\in \mathcal{A}$ and let $i$ be a player which dictates $A$'s makespan, i.e. $i\in \arg\max_{k\in [n]}{\Sigma_{j\in A_k{}}{t_{k}^{j}}}$. The allocation $A$ is said to be $\beta$--unbalanced with respect to player $i$ if it holds that:

\begin{equation}\label{dummy_app}
\beta=\max_{A'\in \mathcal{A}}{\frac{\Sigma_{j\in A_{i}}t_{i}^{j} - \Sigma_{j\in A'_{i}}t_{i}^{j}}{ms(A',T)}}
\end{equation} where $ms$ is a function that given an allocation and an instance returns its makespan.
From here on we use $A'$ to denote an allocation that achieves the maximum value computed in~\eqref{dummy_app}. 


\begin{claim}[Christodoulou et al. \cite{CKV07}]\label{d_claim} 
Fix a mechanism $M$. Let $T$ be an instance with a $\mathcal{D}$ matrix for which $M$ outputs a $\beta$--unbalanced allocation $A$ (for $\beta>0$) with respect to some player $i$ that dictates $A$'s makespan. Then, $M$ has an approximation ratio of at least $1+\beta$.
\end{claim}


We now sketch the proof of the claim.
Increase the $i$'th player's cost for his dummy job to $\gamma= ms(A',T) - \Sigma_{j\in A'_{i}}{t_{i}^{j}}$. If $M$ has a bounded approximation ratio, then by a weak monotonicity argument  $M$ will yield the same allocation. In this case, the optimal makespan is $ms(A',T)$ which results in $M$ having an approximation ratio of $\frac{\gamma+\Sigma_{j\in A_{i}}t_{i}^{j}}{ms(A',T)}=1+\beta$.


\subsubsection{A $\mathcal{BCD}$ Instance}\label{bcd_instance}

In this section we introduce the matrix $\mathcal{C}$. This matrix allows us to increase the imbalance of certain allocations in addition to the increase guaranteed by Claim~\ref{d_claim} and thus improve our bounds (see Claim~\ref{bcd_claim}). For example the matrix $\mathcal{B}_{new}$ that is defined in the end of Section~\ref{previous_proofs}, combined with only a $\mathcal{D}$ matrix gives a $1+\sqrt{2}$ lower bound using Claim~\ref{d_claim}, but by adding a $\mathcal{C}$ matrix to this instance we can also use Claim~\ref{bcd_claim} and get a lower bound of $2.8019$ (see Section~\ref{warm-up}).


We start by introducing some notation. A job is said to be \emph{trivial} if there is a player whose cost for the job is at most $\varepsilon$, and \emph{non--trivial} otherwise. Note that a trivial job does not meaningfully affect the makespan.
A player is said to be \emph{active} for a job if his cost for this job is finite. Likewise, a player is said to be active for a set of jobs if there is at least one job in it for which he is active. 
Observe that a mechanism with a finite approximation ratio will allocate a job only to a player which is active for it, as long as the optimal makespan is finite.

When we consider an instance with all three matrices $\mathcal{BCD}$, we assume that the sets of active players for the jobs in the $\mathcal{B}$ matrix and for the jobs in the $\mathcal{C}$ matrix has at most one player in common and if such player exists it is the first player.  




\begin{claim}\label{bcd_claim}
Given an instance $T=\mathcal{BCD}$, if a mechanism $M$ allocates all non--trivial jobs in $\mathcal{B}$ to the first player then it has an approximation ratio of at least $\min{\{1+a, 1+\frac{T_B^1+\frac{x(a^{-k}-1)}{(a^{-1}-1)}}{ms(O^{-1},T)}\}}$  where $k$ is the number of jobs in $\mathcal{C}$,  $T_B^1$ is the time it takes the first player to complete all non--trivial jobs in $\mathcal{B}$, $x$ is the time it takes him to complete the first job in $\mathcal{C}$ and $O^{-1}$ is an allocation with the minimal makespan among all the allocations in which the first player does not participate. 
\end{claim}


Recall that the costs in the matrix $\mathcal{C}$ depend on the parameter $a$.
Claim~\ref{bcd_claim} follows from the next lemma, which is a slightly extended version of Kotsoupias and Vidali \cite{KV}.

\begin{lemma}\label{c_lemma}
Fix an instance $T=\mathcal{BCD}$ and a mechanism $M$. Suppose that M allocates all non--trivial jobs in B to the first player and it has an approximation ratio better than 1+a. Then, $M$ also allocates all jobs in $\mathcal{C}$ to the first player. 
\end{lemma}

To see why this lemma holds, suppose toward contradiction that the first player does not get all jobs in $\mathcal{C}$ and denote by $j$ the first such job. Reduce the cost of the first player for all jobs he does get in $\mathcal{C}$ and all non--trivial jobs in $\mathcal{B}$ to 0. Now (by a simple weak monotonicity argument), we have that some $i$'th player for $i>1$ gets the $j$'th job in $\mathcal{C}$, a job for which his cost is larger than the cost of the first player by a factor of at least $a$. It implies that we have an allocation with makespan larger by a factor of at least $a$ than the optimal one, since we reduced the cost of the first player for all jobs with larger costs to 0.
Thus, we have a $\beta$--unbalanced allocation for $\beta \geq a$ and by Claim~\ref{d_claim}, $M$ has an approximation ratio of at least 1+a.

We now show how to derive Claim~\ref{bcd_claim} from Lemma~\ref{c_lemma} and Claim~\ref{d_claim}. Given an instance $T$ and a mechanism $M$ as in the claim, we can apply Lemma~\ref{c_lemma} and have that if $M$ has a better approximation ratio than $1+a$ then it also allocates all jobs in $\mathcal{C}$ to the first player. That is, we have a $\beta$--unbalanced allocation in $T$ for $\beta=\frac{T_B^1+\frac{x(a^{-k}-1)}{(a^{-1}-1)}}{ms(O^{-1},T)}$. By Claim~\ref{d_claim}, $M$ has an approximation ratio of at least $1+\beta$. Therefore, $M$ has an approximation ratio of at least $\min{\{1+a, 1+\frac{T_B^1+\frac{x(a^{-k}-1)}{(a^{-1}-1)}}{ms(O^{-1},T)}\}}$.

\subsubsection{Re--deriving Previous Lower Bounds}\label{previous_proofs}
We now show how to obtain the previous lower bounds using the $\mathcal {B,C,D}$ point of view.
\vspace{0.01\linewidth}

In \cite{NR99}, Nisan and Ronen introduced the problem and proved a lower bound of 2 for two players. We discuss a ``modern" view of their proof, in which the main instance has one job with the same cost for both players and two dummy jobs, one for each player. I.e., they start with an instance that consists of the concatenation of two matrices, $\mathcal{B}$ (denoted by $\mathcal{B}_{NR}$ below) and $\mathcal{D}$ a matrix of dummy jobs for the two players (there is no $\mathcal{C}$ matrix). The mechanism allocates the non--dummy item to one player, thus this allocation is $1$--unbalanced (by considering the allocation that allocates the non--dummy item to the other player). By Claim \ref{d_claim} the approximation ratio is no better than $2$. Specifically, we get a lower bound of 2 by increasing the cost of the player which gets the job in $\mathcal{B}_{NR}$ for his dummy job to 1.   
$$
\mathcal{B}_{NR}=\begin{pmatrix}
1\\
1
\end{pmatrix}
$$  

Christodoulou et al. proved a lower bound of $1+\sqrt{2}$ for 3 players and 5 jobs in \cite{CKV07}. They introduced the concept of dummy--jobs i.e., the matrix $\mathcal D$ and a revised $\mathcal{B}$ matrix, denoted by $\mathcal{B}_{CKV}$ (depicted below). In their instance there were only 2 parts, $\mathcal{B}$ and $\mathcal{D}$. 
Their proof starts from analyzing the instance $\mathcal {B}_{CKV}\mathcal{D}$ and continues by considering other $\mathcal{BD}$ instances with different $\mathcal{B}$ matrices. In their proof they used a similar argument to Claim~\ref{d_claim}.

$$
\mathcal{B}_{CKV}=\begin{pmatrix}
1 & 1\\
1 & 1\\
1 & 1
\end{pmatrix}
$$

Then, Koutsoupias and Vidali proved in \cite{KV}  a lower bound of $1+\phi$ where the number of players (and jobs) approaches $\infty$. 
Their instance has 2 parts, a matrix of dummy--jobs for all players $\mathcal{D}$ and a matrix which is essentially $\mathcal{C}$ (denoted by $\mathcal{C}_{KV}$ below). They were the first to use a $\mathcal{C}$--like matrix.

We get this lower bound by applying Claim~\ref{bcd_claim} where $\mathcal{B}$ is empty and the $\mathcal{C}$ matrix is the $\mathcal{C}_{KV}$ matrix. Since $\mathcal{B}$ is empty every mechanism allocates all non--trivial jobs in $\mathcal{B}$ to the first player and thus every mechanism has an approximation ratio of at least $\min{\{1+a, 1+\frac{1}{a-1}}\}$ when the number of jobs in $\mathcal{C}$ approaches $\infty$. For $a=\phi$ the approximation ratio in both cases is tight and equals $1+\phi$. 

$$
        \mathcal{C}_{KV}=
\begin{tikzpicture}[baseline=(m1-3-1.base)]
\matrix [matrix of math nodes,left delimiter=(,right delimiter=)] (m1)
{
\frac{1}{a}  &  \frac{1}{a^2}  &  \dots & \frac{1}{a^i} & \dots &\frac{1}{a^n}\\
1  &  \infty &\dots & \dots &\dots &\dots\\
\infty  &  \frac{1}{a} &\infty & \dots & \dots & \dots\\
\vdots &\vdots &\ddots &\vdots &\vdots &\vdots\\
\infty & \dots &\dots &\frac{1}{a^{i-1}} &\infty &\dots\\
\vdots &\vdots &\vdots &\vdots &\ddots &\vdots\\
\infty & \dots &\dots &\dots &\dots &\frac{1}{a^{n-1}}\\
};
\end{tikzpicture}
$$

Recently, Giannakopoulos et al. improved the lower bound to $2.755$ in \cite{GHP20}. 
They were the first to realize that in fact the matrices $\mathcal{CD}$ can be used to get better lower bounds starting from certain unbalanced allocations comparing to instances that use only the matrix $\mathcal{D}$ (Claim~\ref{bcd_claim}). Due to issues with symmetry breaking in the instance $\mathcal{B}_{CKV}$, the matrix they used was not exactly a  $\mathcal{C}$ matrix. However, building on their idea of combining all three matrices and using the $\mathcal{B}_{new}$ matrix
(together with $\mathcal{C}$ and $\mathcal{D}$ matrices) yields a lower bound of $2.8019$, this is shown in Section~\ref{warm-up}.

$$
\mathcal{B}_{new}=\begin{pmatrix}
b_1 & \frac{2}{a} & \frac{2}{a} \\
1   & \varepsilon & \infty      \\
1   & \infty      & \varepsilon \\
\end{pmatrix}
$$

\subsection{Warm--Up: a 2.8019 Lower Bound}\label{warm-up}

The 2.8019 lower bound is in fact a special case of the construction used in the $3-\delta$ lower bound's proof where there is only one block, i.e, where the main instance is $\mathcal{B}_{new}\mathcal{CD}$. Hence, here we only provide a sketch of the proof for this special case which in particular demonstrates the use of Claim~\ref{bcd_claim}.

First, we will state Lemma~\ref{2.8_lemma} and show how to derive the $2.8019$ lower bound from it and Claim~\ref{bcd_instance}. Afterwards we will provide a sketch of the proof for the lemma.  

\begin{lemma}\label{2.8_lemma}
If a mechanism $M$ has an approximation ratio better than $1+a$ (for $a<2$) then it allocates to the first player all the non--trivial jobs of the $\mathcal{B}$ matrix in at least one of the instances $\mathcal{B}_{new}\mathcal{C}\mathcal{D},\mathcal{B}^1_{new}\mathcal{C}\mathcal{D},\mathcal{B}^2_{new}\mathcal{C}\mathcal{D}$, where:
$$
\mathcal{B}_{new}=\begin{pmatrix}
\frac{2}{a} & \frac{2}{a} & \frac{2}{a} \\
1   & \varepsilon & \infty      \\
1   & \infty      & \varepsilon \\
\end{pmatrix}
\mathcal{B}^1_{new}=\begin{pmatrix}
\frac{1}{a} & \frac{1}{a} & \frac{2}{a} \\
1   & 1 & \infty      \\
1   & \infty      & \varepsilon \\
\end{pmatrix}
\mathcal{B}^2_{new}=\begin{pmatrix}
\frac{1}{a} & \frac{2}{a} & \frac{1}{a} \\
1   & \varepsilon & \infty      \\
1   & \infty      & 1 \\
\end{pmatrix}
$$ 
\end{lemma}

This lemma guarantees that for each mechanism $M$ with approximation ratio better than $1+a$  the conditions of Claim~\ref{bcd_claim}
holds in at least one the three instances: $\mathcal{B}_{new}\mathcal{C}\mathcal{D},\mathcal{B}^1_{new}\mathcal{C}\mathcal{D},\mathcal{B}^2_{new}\mathcal{C}\mathcal{D}$.
Observe that in each of these instances the time it takes the first player to complete all non--trivial jobs is $\frac{2}{a}$ and the minimal makespan of an allocation in which the first player does not participate is 1. 
Therefore, each mechanism $M$ has an approximation ratio of at least 
$$\min\{{1+a, 1+\frac{\frac{2}{a}+\frac{\frac{1}{a^2}}{1-a^{-1}}}{1}}\}=\min{\{1+a,1+\frac{2}{a}+\frac{1}{a(a-1)}\}}$$ when the number of jobs in $\mathcal{C}$ approaches $\infty$. By choosing $a=1.8019$ we get that the approximation ratio is at least $2.8019$.

This lemma is proved by considering the possible allocations of the first job in $\mathcal{B}_{new}$. If the first player gets it, the claim is proved. Otherwise, the second or third player gets it, suppose that it is the second player (the analysis is similar in the other case). 

We now show that the second player also gets the second job in $\mathcal{B}_{new}$. Suppose not and then reduce his cost for the first job to 0. By a standard weak monotinicity argument (Lemma~\ref{lemma1}) he will still not get the second job. Then, the first or third player gets the second job and this allocation has a makespan of at least $\frac{2}{a}$ whereas the optimal makespan is $\frac{1}{a}$ (because of the jobs in the $\mathcal{C}$ matrix). Using a dummy job we can get an approximation ratio of 3.

Thus, we can assume that $M$ allocates to the second player the first two jobs in $\mathcal{B}_{new}$. Increase the second player's cost for the second job to 1 and consider the two possible cases. If the second player keeps the second job then by a weak monotinicity argument (Lemma~\ref{lemma4}) he also keeps the first job. This allocation is $a$--unbalanced and using Claim~\ref{d_claim} we get an approximation ratio of $1+a$.
Otherwise, the first player gets the second job. Now, reduce the cost of the first player for the first two jobs to $\frac{1}{a}$, which yields the $\mathcal{B}^1_{new}$ instance. 

By a weak monotinicity argument (Lemma~\ref{lemma_2}) the first player gets at least one of the first two jobs in $\mathcal{B}^1_{new}$. If he gets both the claim is proved. Thus, suppose the first player gets only one and that is the first job (if he gets only the second job the analysis is similar). Reduce the cost of the first player for the first job to 0 and by a weak monotinicity argument (Lemma ~\ref{lemma1}) the allocation remains the same. The optimal makespan of the new instance is $\frac{1}{a}$ (achieved when the first player gets the first two jobs) but the allocation has a makespan of at least 1. Then, this allocation is $a$--unbalanced and using Claim~\ref{d_claim} we get approximation ratio of $1+a$. This concludes the proof of the lemma.

\subsection{The $3-\delta$ Lower Bound}\label{intuition_3}

In Section~\ref{main_setion} a lower bound of $3-\delta$ is proved where the main instance consists of all three matrices. The $\mathcal{C}$ matrix is practically the $\mathcal{C}$ matrix in Figure~\ref{BCD} (the exact matrix is described in Section ~\ref{main_instance}) and $\mathcal{D}$ is a matrix of dummy--jobs for all players.
The $\mathcal{B}$ matrix is a matrix of $r$ blocks of $\mathcal{B}_{new}$ each with costs smaller than the one before, this matrix is denoted $\mathcal{B}_r$ and is depicted below (in the depiction the first block is colored by red, the second by yellow, the $i$'th by green and the $r$'th by blue).

\begin{equation*}
\mathcal{B}_r=
\begin{tikzpicture}[baseline=(m-3-1.base)]
\matrix [matrix of math nodes,left delimiter=(,right delimiter=)] (m)
{
b_1  &   \frac{2}{a}  &  \frac{2}{a}  &  b_2  &   \frac{2}{a^2}  &  \frac{2}{a^2}  &  \dots  &  b_i  &   \frac{2}{a^i}  &  \frac{2}{a^i}  &  \dots  &  b_r  &   \frac{2}{a^r}  &  \frac{2}{a^r}  \\
1  &   \varepsilon  &  \infty  &  \dots  &   \dots  &  \dots  &  \dots  &  \dots  &   \dots  &  \dots  &  \dots  &  \dots  &   \dots  &  \dots  \\
1  &   \infty  &  \varepsilon  &  \dots  &   \dots  &  \dots  &  \dots  &  \dots  &   \dots  &  \dots  &  \dots  &  \dots  &   \dots  &  \dots  \\
\infty  &   \infty  &  \infty  &  \frac{1}{a}  &   \varepsilon  &  \infty  &  \dots  &  \dots  &   \dots  &  \dots  &  \dots  &  \dots  &   \dots  &  \dots\\
\infty  &   \infty  &  \infty  &  \frac{1}{a}  &   \infty  &  \varepsilon  &  \dots  &  \dots  &   \dots  &  \dots  &  \dots  &  \dots  &   \dots  &  \dots  \\
\vdots  &   \vdots  &  \vdots  &  \vdots  &   \vdots  &  \vdots  &  \ddots  &  \vdots  &   \vdots  &  \vdots  &  \vdots  &  \vdots  &   \vdots  &  \vdots  \\
\infty  &   \dots  &  \dots  &  \dots  &   \dots  &  \dots  &  \dots  &  \frac{1}{a^{i-1}}  &   \varepsilon  &  \infty  &  \dots  &  \dots  &   \dots  &  \dots\\
\infty  &   \dots  &  \dots  &  \dots  &   \dots  &  \dots  &  \dots  &  \frac{1}{a^{i-1}}  &   \infty  &  \varepsilon  &  \dots  &  \dots  &   \dots  &  \dots\\
\vdots  &   \vdots  &  \vdots  &  \vdots  &   \vdots  &  \vdots  &  \vdots  &  \vdots  &   \vdots  &  \vdots  &  \ddots  &  \vdots  &   \vdots  &  \vdots\\
\infty  &   \dots  &  \dots  &  \dots  &   \dots  &  \dots  &  \dots  &  \dots  &   \dots  &  \dots  &  \dots  &  \frac{1}{a^{r-1}}  &   \varepsilon  &  \infty\\
\infty  &   \dots  &  \dots  &  \dots  &   \dots  &  \dots  &  \dots  &  \dots  &   \dots  &  \dots  &  \dots  &  \frac{1}{a^{r-1}}  &   \infty  &  \varepsilon\\
};
\draw[rounded corners,ultra thick, draw=black, fill=red, opacity=0.3] (m-3-1.south west) rectangle (m-1-3.north east);
\draw[rounded corners,ultra thick, draw=black, fill=yellow, opacity=0.3] (m-5-4.south west) rectangle (m-1-6.north east);  
\draw[rounded corners,ultra thick, draw=black, fill=green, opacity=0.3] (m-8-8.south west) rectangle (m-1-10.north east); 
\draw[rounded corners,ultra thick, draw=black, fill=blue, opacity=0.3] (m-11-12.south west) rectangle (m-1-14.north east); 
\end{tikzpicture}
\end{equation*}

This lower bound is an improvement of the 2.8019 lower bound (Section~\ref{warm-up}), the improvement is achieved by using multiple blocks in the $\mathcal{B}$ matrix instead of one.
Now, we give some intuition for the reason why multiple blocks allows us to improve the lower bound shown in Section~\ref{warm-up}.


Consider the allocation of the first job in $\mathcal{B}_{new}$. There are two possible cases. In the first case, the first player gets this job, therefore we would want the value $b_1$ to be relatively large. In the second case, some other player gets the first job in $\mathcal{B}_{new}$.
Then using weak monotonicity arguments we change the original matrix to one with two non-trivial jobs such that the first player gets both jobs where his costs for these jobs are higher as the value of $b_1$ is smaller. Then, in this case, we would want $b_1$ to be relatively small. The value of $b_1$ is chosen  such that it balances between the two opposite constraints. 

Having multiple blocks relaxes the two constraints.
One of them has more weight in the first few blocks and the other in the last few blocks. This is why the ratio $\frac{b_i}{a^{-i}}$ (the relative value of $b_i$) changes throughout the blocks, starting with a high ratio that decreases along the blocks.

\section{Preliminaries}

There are $n$ machines, $m$ tasks. Denote by $t_{i}^{j}$ the time it takes machine $i$ to process job $j$. The cost of machine $i$, denoted by $t_{i}=(t_{i}^{1},...t_{i}^{m})$, is its private information. A time--processing matrix $T\in {\mathbb{R}}^{n\times m}$ is a matrix where the $i$'th row is $t_{i}$. Let $x=(x_{1},...,x_{n})$ denote an allocation of tasks to machines, where $x_{i}$ is the set of tasks allocated to machine $i$ and $x_i^j $ equal 1 if machine $i$ gets job $j$ in $x$ and 0 otherwise. A valid allocation allocates each task to exactly one machine. Let $A$ be the set of all valid allocations.
We are interested in truthful mechanisms, which are mechanisms where the dominant strategy of each agent (machine) is to reveal his true type $t_{i}$.
A truthful mechanism is a tuple $M=(f,P)$ that consists of an allocation function $f:{\mathbb{R}}^{n\times m}\to A$ and a payment scheme $p:{\mathbb{R}}^{n\times m}\to R^{n}$. The objective is to minimize the makespan, which is given by $\max_{i\in[n]}\Sigma_{j\in x_{i}}t_{i}^{j}$.
Each machine is controlled by a selfish agent whose goal is to maximize his utility function: $u_{i}(x,p)=p_{i}-\Sigma_{j\in x_{i}}t_{i}^{j}$ (the mechanism pays the agents in order to incentivize them to perform the tasks).

A known characterization of a truthful mechanism is that its social choice function is weakly--monotone \cite{BC06}. In the setting of unrelated machine scheduling a mechanism is truthful if and only if it has a weakly--monotone allocation algorithm.
An allocation algorithm $f:{\mathbb{R}}^{n\times m}\to A$ is \emph{weakly monotone} if for every agent $i$, every $t_{-i}$ and every $T=(t_i,t_{-i}), T'=(t'_i,t_{-i})$ it holds that:
\begin{align*}
\sum_{j\in{[m]}}{({t_i}^j-{t'_i}^j)\cdot({x_i}^j-{x'_i}^j)}\leq 0
\end{align*}
 where $ x=f(T), x'=f(T')$.    

We will use $\infty$ to denote very large values, and $\varepsilon$ to denote values that are as small as we wish.
A dummy job for player $i$ is a job which takes player $i$ some finite time to execute, while for all the other players it takes infinite time. That is, every mechanism which achieves a finite approximation must allocate this job to player $i$.

Several properties which follow from the weak monotonicity characterization are given in the lemmas below. After each lemma we provide an example to illustrate it. Similar lemmas are standard in the related literature.

\begin{lemma}\label{lemma1}
 Let $M$ be a truthful mechanism with an allocation function $f$, and let
$T,T'\in\mathbb{R}^{n\times m}$
be time--processing matrices that differ only on player $i$, i.e., $T_{-i}=T'_{-i}$ and $x=f(T), x'=f(T')$. Let $F_1=\{j\,|\,x^j_i=1 \land {t_i}^j>{t'_i}^j\}$, $F_2=\{k\,|\,x^k_i=0 \land {t_i}^k<{t'_i}^k\}$, and $F_3=[m]\setminus (F_1\cup F_2)$. Suppose that for every $q\in F_3$, it holds that $t^q_i=t'^q_i$. Then, for every $r\in(F_1\cup F_2)$, it holds that $x^r_i=x'^r_i$.
\end{lemma}

\begin{proof}[Proof of Lemma \ref{lemma1}]
By weak monotonicity:
\begin{align*}
    0\geq &\sum_{p\in{[m]}}{({t_i}^p -{t'_i}^p) \cdot ({x_i}^p {-x'_i}^p)}\\
    &= \sum_{j\in{F_1}}{({t_i}^j-{t'_i}^j)\cdot({x_i}^j-{x'_i}^j)} + \sum_{k\in{F_2}}{({t_i}^k-{t'_i}^k)\cdot({x_i}^k-{x'_i}^k)}\\
     &= \sum_{j\in{F_1}} {\underbrace{({t_i}^j-{t'_i}^j)}_{ >0} \cdot  \underbrace{(1-{x'_i}^j)}_{\geq 0}}
     +\sum_{k\in{F_2}}{\underbrace{({t_i}^k-{t'_i}^k)}_{< 0}\cdot\underbrace{(0-{x'_i}^k)}_{\leq 0}}\\
\end{align*}
In order for the weak monotonicity inequality to hold, each term in the summation must be equal to 0, i.e., for every job $r$ such that $r\in (F_1\cup F_2)$, it holds that ${x'^i}_r=x^i_r$ , and for any other job $k\in F_3$, ${x'^i}_k=0$.
\end{proof}

When using Lemma \ref{lemma1}, if not stated otherwise, $F_1$ will be the set of all the jobs that the $i$'th player gets in $x$ and $F_2$ will be the set of all the jobs he does not get in $x$.

\begin{example}[an example of Lemma \ref{lemma1}]
\label{example_1}
Consider the instances $G_1$ with the allocation indicated by stars and the instance $G_2$ (given below). Applying Lemma \ref{lemma1} on these instances where $T=G_1, \, T'=G_2,\,i=2, \,F_1=\{3\}, \, F_2=\{1\}$ and $ \,F_3=\{2\}$ gives us that in the instance $G_2$ the second player gets the third job but not the first one. 
\
\[
\underbrace{\begin{pmatrix}
1^*  &2  &3\\
2  &1^*  &3^*
\end{pmatrix}}_{=G_1}
\xrightarrow{}
\quad
\underbrace{\begin{pmatrix}
1^*  &2  &3\\
3  &1  &2^*
\end{pmatrix}}_{=G_2}
\]
\end{example}

\begin{lemma}\label{lemma_2}
Let $M$ be a truthful mechanism with a social choice function $f$, let 
$T,T'\in\mathbb{R}^{n\times m}$
be time--processing matrices that differ only on player $i$, i.e., $T_{-i}=T'_{-i}$, and let $x=f(T), x'=f(T')$. Let $j$ be a job that player $i$ gets in $x$ where ${t_i}^j>{t'_i}^j$, and let $k$ be another job such that ${t_i}^k>{t'_i}^k$. Suppose that for any other job $l$, it holds that ${t_i}^l={t'_i}^l$ . Then, it follows that either ${x'_i}^j=1$ or ${x'_i}^k=1$ or both. If in addition it holds that $({t_i}^k-{t'_i}^k)<({t_i}^j>{t'_i}^j)$ then ${x'_i}^j=1$.
\end{lemma}

\begin{proof}[Proof of Lemma \ref{lemma_2}]
By weak monotonicity:
\begin{equation}\label{WMON_ineq}
\begin{split}
    0 \geq &\sum_{r\in{[m]}}{({t_i}^r -{t'_i}^r) \cdot ({x_i}^r {-x'_i}^r)}\\
    &= ({t_i}^j -{t'_i}^j) \cdot ({x_i}^j {-x'_i}^j)+({t_i}^k -{t'_i}^k) \cdot ({x_i}^k {-x'_i}^k)\\
    &=\underbrace{({t_i}^j -{t'_i}^j)}_{>0} \cdot (1 {-x'_i}^j)+\underbrace{({t_i}^k -{t'_i}^k)}_{>0} \cdot ({x_i}^k {-x'_i}^k)
\end{split}
\end{equation}

Suppose not. Then, ${x'_i}^j = {x'_i}^k=0$, and we have that:
\begin{align*}
    & \sum_{r\in{[m]}}{({t_i}^r -{t'_i}^r) \cdot ({x_i}^r {-x'_i}^r)}\\
    &=\underbrace{({t_i}^j -{t'_i}^j)}_{>0}+\underbrace{({t_i}^k -{t'_i}^k)}_{>0} \cdot ({x_i}^k)
\end{align*}

and so the inequality does not hold, therefore either ${x'_i}^j = 1$ or ${x'_i}^k = 1$.
If we also have that $({t_i}^k-{t'_i}^k)<({t_i}^j>{t'_i}^j)$ then in order for inequality~\eqref{WMON_ineq} to hold it must be the case that ${x'_i}^j = 1$.
\end{proof}

\begin{example}[an example of Lemma \ref{lemma_2}]
\label{example_2}
Consider the instances $H_1$ with the allocation indicated by stars and the instance $H_2$  (given below). Applying Lemma \ref{lemma_2} on these instances where $T=H_1, \, T'=H_2,\, i=2, \,j=2$ and $ \,k=1$ gives us that in the instance $H_2$ the second player gets at least one of the first two jobs. 
\
\[
\underbrace{\begin{pmatrix}
1^*  &2  &3\\
2  &1^*  &3^*
\end{pmatrix}}_{=H_1}
\xrightarrow{}
\quad
\underbrace{\begin{pmatrix}
1  &2  &3\\
1  &0  &3
\end{pmatrix}}_{=H_2}
\]
\end{example}

\begin{lemma}\label{lemma3}
Let $M$ be a truthful mechanism with a finite approximation ratio, its social choice function is denoted by $f$. Let $i$ be a player with a dummy job $j$, let $T,T'\in\mathbb{R}^{n\times m}$ be time--processing matrices that differ only on player $i$, i.e., $T_{-i}=T'_{-i}$, and let $x=f(T)$. Let $F_1=\{r\neq j\,|\,x^r_i=1 \land {t_i}^r>{t'_i}^r\}$, $F_2=\{k\,|\,x^k_i=0 \land {t_i}^k<{t'_i}^k\}$, and $F_3=[m]\setminus (F_1\cup F_2\cup \{j\})$. Suppose that for every $q\in F_3$, it holds that $t^q_i=t'^q_i$.  Then for $p\in (F_1\cup F_2\cup \{j\})$, it holds that $x^p_i=x'^p_i$, where $x'=f(T')$.
\end{lemma}

\begin{proof}[Proof of Lemma \ref{lemma3}]
Observe that if the mechanism has a finite approximation ratio then it must allocate job $j$ to player $i$. Thus, we have that ${x_i}^j=1$ and ${x'_i}^j=1$ .
By weak monotonicity:
\begin{align*}
    0 \geq &\sum_{s\in{[m]}}{({t_i}^s -{t'_i}^s) \cdot ({x_i}^s {-x'_i}^s)}\\
    &= \sum_{r\in{F_1}}{({t_i}^r-{t'_i}^r)\cdot({x_i}^r-{x'_i}^r)} + \sum_{k\in{F_2}}{({t_i}^k-{t'_i}^k)\cdot({x_i}^k-{x'_i}^k)}+({t_i}^j-{t'_i}^j)\cdot \underbrace{({x_i}^j-{x'_i}^j)}_{=0}\\
    &= \sum_{r\in{F_1}}{\underbrace{({t_i}^r-{t'_i}^r)}_{>0}\cdot\underbrace{(1-{x'_i}^r)}_{\geq 0}}+ \sum_{k\in{F_2}}{\underbrace{({t_i}^k-{t'_i}^k)}_{<0}\cdot\underbrace{(0-{x'_i}^k)}_{\leq 0}}
\end{align*}
and for the inequality to hold it must be that for every $r \in F_1$, it holds that $x^r_i=x'^r_i=1$ and for every $k \in F_2$, it holds that $x^k_i=x'^k_i=0$. As explained before, since the mechanism has a finite approximation ratio it must be that ${x_i}^j={x'_i}^j=1$.
\end{proof}

When using Lemma \ref{lemma3}, if not stated otherwise, $F_1$ will be the set of all the jobs that the $i$'th player gets in $x$ and $F_2$ will be the set of all the jobs he does not get in $x$.

\begin{example}[an example of Lemma \ref{lemma3}]
\label{example_3}
Consider the instance $I_1$  with the allocation indicated by stars and the instance $I_2$ (given below). Applying Lemma \ref{lemma3} on these instances where $T=I_1, \, T'=I_2,\, i=2, \, j=4, \, F_1=\{3\}, \,F_2=\{1\}$ and $ \,F_3=\{2\}$ gives us that in instance $I_2$ the second player gets the third and forth jobs but not the first job. 
\
\[
\underbrace{\begin{pmatrix}
1^*  &1^* &2  &\infty\\
1    &2   &1^*  &1^*
\end{pmatrix}}_{=I_1}
\xrightarrow{}
\quad
\underbrace{\begin{pmatrix}
1^*  &1^* &2  &\infty\\
2    &2   &{\frac{1}{2}}^*  &3^*
\end{pmatrix}}_{=I_2}
\]
\end{example}

\begin{lemma}\label{lemma4}
Let $M$ be a truthful mechanism with a social choice function $f$, let 
$T,T'\in\mathbb{R}^{n\times m}$
be time--processing matrices that differ only on player $i$, i.e., $T_{-i}=T'_{-i}$, and let $x=f(T), x'=f(T')$.
Let $j_1,j_2$ be two jobs such that ${x_i}^{j_1}={x_i}^{j_2}=1$, ${t_i}^{j_1}>{t'_i}^{j_1}$ and ${t_i}^{j_2}<{t'_i}^{j_2}$. 
Suppose that for any other job $j\neq j_1,j_2$, it holds that  ${t_i}^{j}={t'_i}^j$.  Then, if ${x'_i}^{j_2}=1$, then also ${x'_i}^{j_1}=1$.
\end{lemma}

\begin{proof} [Proof of Lemma \ref{lemma4}]
Assume that ${x'_i}^{j_2}=1$. By weak monotonicity:
\begin{align*}
    0 \geq &\sum_{s\in{[m]}}{({t_i}^s -{t'_i}^s) \cdot ({x_i}^s {-x'_i}^s)}\\
    &= ({t_i}^{j_1}-{t'_i}^{j_1})\cdot({x_i}^{j_1}-{x'_i}^{j_1})+ ({t_i}^{j_2}-{t'_i}^{j_2})\cdot({x_i}^{j_2}-{x'_i}^{j_2})\\
        &= ({t_i}^{j_1}-{t'_i}^{j_1})\cdot(1-{x'_i}^{j_1})+ ({t_i}^{j_2}-{t'_i}^{j_2})\cdot(1-1)\\
    &= \underbrace{({t_i}^{j_1}-{t'_i}^{j_1})}_{>0}\cdot\underbrace{(1-{x'_i}^{j_1})}_{\geq 0}\\
\end{align*}
In order for the inequality to hold it must be the case that ${x'_i}^{j_1}=1$.
\end{proof}

\begin{example}[an example of Lemma \ref{lemma4}]
\label{example_4}
Consider the instance $K_1$ with the allocation indicated by stars and the instance $K_2$ (given below). Applying Lemma \ref{lemma4} on these instances where $T=K_1, \, T'=K_2,\, i=2, \, j_1=1$ and $ \,j_2=2$ gives us that in the instance $K_2$ if the second player gets the second job then he must also get the first job. 
\
\[
\underbrace{\begin{pmatrix}
1      &2     &3^*  \\
1^*    &0^*   &3 
\end{pmatrix}}_{=K_1}
\xrightarrow{}
\quad
\underbrace{\begin{pmatrix}
1      &2     &3  \\
\frac{3}{4}    &1   &3 
\end{pmatrix}}_{=K_2}
\]
\end{example}

When applying lemmas \ref{lemma1}, \ref{lemma_2}, \ref{lemma3}, \ref{lemma4} we will often increase or decrease only some of the costs, and not all of them, in order to make the proof clearer. The convention will be that the other values are increased or decreased by a small amount, as was done in \cite{KV}.

When visualizing instances using matrices, a blue star (\textcolor{blue}{*}) will be used to indicate the allocation of the mechanism $M$ in this instance, and a red at (\textcolor{red}{@}) will be used to indicate an optimal allocation in this instance. 
Moreover, note that sometimes the allocation of the mechanism $M$ will be partial (namely, some jobs will not be visually assigned a player, since which player gets the job is irrelevant).

\section{A Lower Bound of $3-\delta$} \label{main_setion}

\begin{theorem} \label{main_thm}
For every $0<\delta<2-\sqrt{2}$, there exists a number $n\in \mathbb{N}$ such that every truthful mechanism for $n$ machines and $\frac{7(n-1)}{3}$ jobs cannot guarantee an approximation ratio better than $3-\delta$. 
\end{theorem}

We prove the following equivalent formulation: for every $\sqrt{2}<a<2$ there exists a number $n\in \mathbb{N}$ such that every truthful mechanism for $n$ machines and $\frac{7(n-1)}{3}$ jobs cannot guarantee an approximation ratio better than $1+a$ (where $a= 2-\delta$).

Section ~\ref{preparations} introduces basic building blocks that will be used in the proof of the theorem and the proof itself appears in Section ~\ref{proof}.
\
\subsection{Preparations for the Proof} \label{preparations}

In this section we introduce some notation, present the main instance (Section ~\ref{main_instance}) and define two transitions  we use in the proof of the theorem (Section ~\ref{e1_e2}).
\par
A job is said to be \emph{trivial} if there is a player whose cost for the job is either zero or $\varepsilon$, and \emph{non--trivial} otherwise. Note that, a trivial job does not meaningfully affect the makespan.
A player is said to be \emph{active} for a job if his cost for this job is finite. Likewise, a player is said to be active for a set of jobs if there is at least one job in it for which he is active. 
Observe that a mechanism with a finite approximation ratio will allocate a job only to a player which is active for it, in the case that the optimal makespan is finite.


\subsubsection{The Main instance $\mathcal{A}_n$} \label{main_instance}
Our instance is defined only for $n = 3r +1$ machines, where $r \in \mathbb{N}$.
Fix a number $a<2$, and let $\mathcal{A}_n= \left[
\begin{array}{c;{2pt/2pt}c;{2pt/2pt}c}
     \mathcal{B}_r & \mathcal{C}_r & \mathcal{D}_r  \\
\end{array}
\right]
$ where \rotatebox{90}{\Kutline} denotes the concatenation of two time processing matrices of the same height (i.e., with the same number of players).  In the instance $\mathcal{A}_n$ there are $n$ players and $ \frac{7(n-1)}{3}=7r$ jobs, where $\mathcal{B}_r, \, \mathcal{C}_r, $ and $\mathcal{D}_r$ consist of $3r$, $r$, and $3r+1$ jobs, respectively.

\begin{center}
\begin{equation*}
\begin{array}{c;{2pt/2pt}c;{2pt/2pt}c}
     \mathcal{B}_r & \mathcal{C}_r & \mathcal{D}_r =  \\
\end{array}
\begin{tikzpicture}[baseline=(m-3-1.base)]
\matrix [matrix of math nodes,left delimiter=(,right delimiter=)] (m)
{
b_1  &   \frac{2}{a}  &  \frac{2}{a}  &  b_2  &   \frac{2}{a^2}  &  \frac{2}{a^2}  &  \dots  &  b_i  &   \frac{2}{a^i}  &  \frac{2}{a^i}  &  \dots  &  b_r  &   \frac{2}{a^r}  &  \frac{2}{a^r}  &  \frac{1}{a^{r+1}}  &  \dots  &  \frac{1}{a^{2r}}  &\dots\\
1  &   \varepsilon  &  \infty  &  \dots  &   \dots  &  \dots  &  \dots  &  \dots  &   \dots  &  \dots  &  \dots  &  \dots  &   \dots  &  \dots  &  \dots  &   \dots  &  \dots  & \dots\\
1  &   \infty  &  \varepsilon  &  \dots  &   \dots  &  \dots  &  \dots  &  \dots  &   \dots  &  \dots  &  \dots  &  \dots  &   \dots  &  \dots  &  \dots  &   \dots  &  \dots  &  \dots\\
\infty  &   \infty  &  \infty  &  \frac{1}{a}  &   \varepsilon  &  \infty  &  \dots  &  \dots  &   \dots  &  \dots  &  \dots  &  \dots  &   \dots  &  \dots  &  \dots  &   \dots  &  \dots  &\dots\\
\infty  &   \infty  &  \infty  &  \frac{1}{a}  &   \infty  &  \varepsilon  &  \dots  &  \dots  &   \dots  &  \dots  &  \dots  &  \dots  &   \dots  &  \dots  &  \dots  &   \dots  &  \dots  &\dots\\
\vdots  &   \vdots  &  \vdots  &  \vdots  &   \vdots  &  \vdots  &  \ddots  &  \vdots  &   \vdots  &  \vdots  &  \vdots  &  \vdots  &   \vdots  &  \vdots  &  \vdots  &   \vdots  &  \vdots  &\dots\\
\infty  &   \dots  &  \dots  &  \dots  &   \dots  &  \dots  &  \dots  &  \frac{1}{a^{i-1}}  &   \varepsilon  &  \infty  &  \dots  &  \dots  &   \dots  &  \dots  &  \dots  &   \dots  &  \dots  &\dots\\
\infty  &   \dots  &  \dots  &  \dots  &   \dots  &  \dots  &  \dots  &  \frac{1}{a^{i-1}}  &   \infty  &  \varepsilon  &  \dots  &  \dots  &   \dots  &  \dots  &  \dots  &   \dots  &  \dots  &\mathcal{D}_r\\
\vdots  &   \vdots  &  \vdots  &  \vdots  &   \vdots  &  \vdots  &  \vdots  &  \vdots  &   \vdots  &  \vdots  &  \ddots  &  \vdots  &   \vdots  &  \vdots  &  \dots  &   \dots  &  \dots  &\dots\\
\infty  &   \dots  &  \dots  &  \dots  &   \dots  &  \dots  &  \dots  &  \dots  &   \dots  &  \dots  &  \dots  &  \frac{1}{a^{r-1}}  &   \varepsilon  &  \infty  &  \dots  &   \dots  &  \dots  &\dots\\
\infty  &   \dots  &  \dots  &  \dots  &   \dots  &  \dots  &  \dots  &  \dots  &   \dots  &  \dots  &  \dots  &  \frac{1}{a^{r-1}}  &   \infty  &  \varepsilon  &  \dots  &   \dots  &  \dots  &\dots \\
\infty  &   \dots  &  \dots  &  \dots  &   \dots  &  \dots  &  \dots  &  \dots  &   \dots  &  \dots  &  \dots  &  \dots  &   \dots  &  \dots  &  \frac{1}{a^{r}}  &   \dots  &  \dots  &\dots  \\
\vdots  &   \vdots  &  \vdots  &  \vdots  &   \vdots  &  \vdots  &  \vdots  &  \vdots  &   \vdots  &  \vdots  &  \vdots  &  \vdots  &   \vdots  &  \vdots  &  \vdots  &   \ddots  &  \vdots  & \dots\\
\infty  &   \dots  &  \dots  &  \dots  &   \dots  &  \dots  &  \dots  &  \dots  &   \dots  &  \dots  &  \dots  &  \dots  &   \dots  &  \dots  &  \infty  &   \dots  &  \frac{1}{a^{2r-1}}  &\dots\\
};
\draw[rounded corners,ultra thick, draw=black, fill=red, opacity=0.3] (m-3-1.south west) rectangle (m-1-3.north east);
\draw[rounded corners,ultra thick, draw=black, fill=yellow, opacity=0.3] (m-5-4.south west) rectangle (m-1-6.north east);  
\draw[rounded corners,ultra thick, draw=black, fill=green, opacity=0.3] (m-8-8.south west) rectangle (m-1-10.north east); 
\draw[rounded corners,ultra thick, draw=black, fill=blue, opacity=0.3] (m-11-12.south west) rectangle (m-1-14.north east); 

\mvline[dashed, black]{m}{15};
\mvline[dashed, black]{m}{18};
\end{tikzpicture}
\end{equation*}
\end{center}

The matrix $\mathcal{B}_r$ consists of $r$ blocks, each block has 3 jobs and 3 active players. The first player will be an active player for all these blocks, whereas the other two active players will be different for each block. More concretely, the only active players for block $i$ will be players $1, 2i,$ and $2i+1$.
The blocks are illustrated as follows. The first (red) rectangle is the first block, the second (yellow) rectangle is the second block, the third (green) rectangle is the $i$'th block, and the forth (blue) rectangle is the $r$'th block. The $i$'th block is denoted by $B_i$.

Every job in $\mathcal{C}_r$ has exactly two active players. For the $i$'th job in $\mathcal{C}_r$, the two players are player $1$ and player $2r+1 + i$. Moreover, the second player's cost is always $a$ times larger than that of the first player.

The matrix $\mathcal{D}_r$ consists of dummy jobs for the $n$ players. The first job is a dummy job for the first player and so on. In other words, the matrix has zeroes on its diagonal and $\infty$ everywhere else. 

The values $b_1,...,b_r$ satisfy $b_i \geq \frac{1}{a^i}$ for every $i$ (Lemma ~\ref{bk_is_atleat}), where the exact values will be specified later.
See Example ~\ref{example_of_main_instance} for a concrete example of $\mathcal{A}_n$ where $a\approx 1.873$ and $r=3, n=10$.

Let $B$ be the set of the first jobs in every block, i.e., $B=\{1+3i \,| \, 0\leq i \leq r-1\}$ and let $C$ be the set of all the jobs in $\mathcal{C}_r$, i.e., $C=\{3r+1,...,4r\}$.

\subsubsection{Transitions $E_1$ and $E_2$} \label{e1_e2}
We define a two--step transition ${E^i_1}$ on a given block $B_i$. 
In the first step, the cost of the $2i$'th player for the second job in the block is increased to  $\frac{1}{a^{i-1}}$, whereas in the second step the first player's cost for the first two jobs in the block is decreased. More specifically, the first player's cost for the first job is decreased to $\frac{1}{a^i}$ (which is indeed a decrease, since $b_i\geq \frac{1}{a^i}$) and his cost for the second job is decreased to $\max\{\frac{2}{a^i}-(b_i-\frac{1}{a^i}+\varepsilon), \frac{1}{a^i}\}$ (this transition is depicted below). The resulting block will be denoted by $B^{E_1}_i$.

\begin{center}
\[
\begin{pmatrix}
b_i                &  \frac{2}{a^i}    &  \frac{2}{a^i}\\
\vdots             &  \vdots           &  \vdots\\
\frac{1}{a^{i-1}}  &  \varepsilon      &\infty\\
\frac{1}{a^{i-1}}  &  \infty           &\varepsilon\\
\end{pmatrix}
\xrightarrow{}
\quad
\begin{pmatrix}
b_i                &  \frac{2}{a^i}          &  \frac{2}{a^i}\\
\vdots             &  \vdots           &  \vdots\\
\frac{1}{a^{i-1}}  &  \frac{1}{a^{i-1}}      &\infty\\
\frac{1}{a^{i-1}}  &  \infty                 &\varepsilon\\
\end{pmatrix}
\xrightarrow{}
\quad
\begin{pmatrix}
\frac{1}{a^i}      & \max\{\frac{2}{a^i}-(b_i-\frac{1}{a^i}+\varepsilon), \frac{1}{a^i}\}  &  \frac{2}{a^i}\\
\vdots             &  \vdots           &  \vdots\\
\frac{1}{a^{i-1}}  &  \frac{1}{a^{i-1}}                                 &\infty\\
\frac{1}{a^{i-1}}  &  \infty                                            &\varepsilon\\
\end{pmatrix}
\]
\end{center}

Similarly, we define a two--step transition ${E^i_2}$ on a block $B_i$. This time, the first step is to increase the $(2i+1)$'th player's cost for the third job in the block to $\frac{1}{a^{i-1}}$, and the second step is to decrease the first player's cost for the third job to $\max\{\frac{2}{a^i}-(b_i-\frac{1}{a^i}+\varepsilon), \frac{1}{a^i}\}$, and (as before) decrease his cost for the first job to $\frac{1}{a^i}$. The resulting block will be denoted by $B^{E_2}_i$ (this transition is depicted below).
For simplicity, we write transition $E_1$ instead of ${E_1}^i$ where the context is clear and we do the same for $E_2$. 
\begin{center}
\[
\begin{pmatrix}
b_i                &  \frac{2}{a^i}    &  \frac{2}{a^i}\\
\vdots             &  \vdots           &  \vdots\\
\frac{1}{a^{i-1}}  &  \varepsilon      &\infty\\
\frac{1}{a^{i-1}}  &  \infty           &\varepsilon\\
\end{pmatrix}
\xrightarrow{}
\quad
\begin{pmatrix}
b_i                &  \frac{2}{a^i}          &  \frac{2}{a^i}\\
\vdots             &  \vdots           &  \vdots\\
\frac{1}{a^{i-1}}  &  \varepsilon      &\infty\\
\frac{1}{a^{i-1}}  &  \infty                 &\frac{1}{a^{i-1}}\\
\end{pmatrix}
\xrightarrow{}
\quad
\begin{pmatrix}
\frac{1}{a^i}      &  \frac{2}{a^i}   &\max\{\frac{2}{a^i}-(b_i-\frac{1}{a^i}+\varepsilon), \frac{1}{a^i}\}\\
\vdots             &  \vdots           &  \vdots\\
\frac{1}{a^{i-1}}  &  \varepsilon                                 &\infty\\
\frac{1}{a^{i-1}}  &  \infty                                            &\frac{1}{a^{i-1}}\\
\end{pmatrix}
\]
\end{center}

Now we can define the following notation. 
For every $1 \leq i\leq r$, a given instance is said to be of the form $\mathcal{A}_{n,i}$ if it is identical to $\mathcal{A}_n$ except that all the jobs in its first $i-1$ blocks are trivial. If in addition the $i$'th block of the instance was transitioned to either $B^{E_1}_i$ or $B^{E_2}_i$, the instance is said to be of the form $A^{E_1}_{n,i}$ or $A^{E_2}_{n,i}$ (or simply $A^E_{n,i}$). For $i=0$, we define $\mathcal{A}_{n,0}=A^E_{n,0}=\mathcal{A}_n$. An example of an instance of the form $A^{E_2}_{n,i}$ for $r=2, n=7$ is given in Figure ~\ref{possible_instances_n=2}.

Next, we state and prove a key lemma regarding the application of $E_1$ and $E_2$ in the proof of the theorem. The lemma is stated and proved for the application of $E_1$ and is similar when applying $E_2$; the needed changes are written in parentheses.   

\begin{lemma}\label{transition_lemma_1}
Consider an instance $A'_{n,i}$ of the form $\mathcal{A}_{n,i}$ for $1 \leq i \leq r$ and let $M$ be a mechanism with approximation ratio better than $1+a$ that allocates the first and second (third) jobs in the block $B_i$ to the $2i$'th ($(2i+1)$'th) player. Then after applying transition $E_1$ ($E_2$) on this block, $M$ allocates the first and second (third) jobs in $B^{E_1}_i$ ($B^{E_2}_i$) to the first player.
\end{lemma}

\begin{proof}[Proof of Lemma \ref{transition_lemma_1}]
Consider the first step of $E_1$ in which we increase the $2i$'th player's cost for the second job in $B_i$ to $\frac{1}{a^{i-1}}$. There are two possible cases. In the first, $M$ allocates the second job in $B_i$ to the $2i$'th player. In the second, $M$ allocates this job to the first player (these are the only two active players for this job). We begin by analyzing the first case.
The $2i$'th player keeps the second job in $B_i$, by Lemma~\ref{lemma4} he must also keep the first job in this block. Now, we increase his dummy job to $\frac{2}{a^i}$ and by Lemma~\ref{lemma3} the allocation of this player remains the same. This causes $M$'s makespan to be $\frac{2}{a^i}+\frac{2}{a^{i-1}}$ whereas the optimal makespan is  $\frac{2}{a^i}$ (for $a\leq 2$) which results in an approximation ratio of $1+a$, this case is depicted below. 

\begin{center}
\begin{tikzpicture}
\matrix [matrix of math nodes,left delimiter=(,right delimiter=)] (m)
{
0& \dots &0   &b_i      &\frac{2}{a_i}^{\textcolor{red}{@}}  &  \frac{2}{a^i} &b_{i+1} &\frac{2}{a^{i+1}} &\frac{2}{a^{i+1}} &\dots & b_r & \frac{2}{a^r} &\frac{2}{a^r}& \dots & \dots & \dots & \dots \\
\vdots   &  \vdots  &  \vdots &\vdots   &  \vdots  &  \vdots &\vdots&  \vdots  &  \vdots &\vdots   &  \vdots  &  \vdots &\vdots &\mathcal{C}_r & \dots &\vdots &\dots\\
\dots &\dots  &\dots  &\frac{1}{a^{i-1}}^{\textcolor{blue}{*}}  &  \frac{1}{a^{i-1}}^{\textcolor{blue}{*}}   &\infty &\dots &\dots&\dots &\dots&\dots &\dots&\dots &\dots & \dots & \frac{2}{a^i}^{\textcolor{blue}{*}{\textcolor{red}{@}}} & \dots\\
\dots & \dots &\dots  &\frac{1}{a^{i-1}}^{\textcolor{red}{@}}   &  \infty  &\varepsilon^{\textcolor{red}{@}}  &\dots &\dots &\dots  & \dots & \dots&\dots &\dots &\dots  & \dots & \dots&\dots\\
\dots &\dots  &\dots&\dots &\dots&\dots  &\frac{1}{a^{i}}^{\textcolor{red}{@}}   &  \varepsilon^{\textcolor{red}{@}}    &\infty  &\dots&\dots &\dots&\dots &\dots & \dots & \dots & \dots\\
\dots &\dots  &\dots&\dots &\dots&\dots  &\frac{1}{a^{i}}  &  \infty   &\varepsilon^{\textcolor{red}{@}}   &\dots&\dots &\dots&\dots &\dots & \dots & \dots & \dots\\
\vdots  &  \vdots  &\vdots  &\vdots &\vdots &\vdots & \vdots &\vdots& \vdots&\ddots& \vdots&\vdots & \vdots &\vdots& \vdots&\vdots& \vdots\\
\dots &\dots  &\dots&\dots &\dots&\dots&\dots &\dots&\dots&\dots &\frac{1}{a^{r}}^{\textcolor{red}{@}}   &  \varepsilon^{\textcolor{red}{@}}    &\infty  &\dots&\dots &\dots &\dots\\
\dots &\dots  &\dots&\dots &\dots&\dots&\dots &\dots&\dots&\dots &\frac{1}{a^{r}}  &  \infty   &\varepsilon^{\textcolor{red}{@}}   &\dots&\dots &\dots &\dots\\
\vdots  &  \vdots  &\vdots  &\vdots &\vdots &\vdots & \vdots &\vdots& \vdots&\vdots& \vdots&\vdots & \vdots &\vdots& \vdots&\vdots& \vdots\\
};
\mvline[dashed, black]{m}{14};
\mvline[dashed, black]{m}{15};
\end{tikzpicture}
\end{center}

Recall that in the second case $M$ allocates the second job in $B_i$ (after the change of first step) to the first player. We now apply the second step of $E_1$ (reducing the first player's cost for the first two jobs in $B^{E_1}_i$) and consider the two possible scenarios.

In the first, $\frac{2}{a^i}-(b_i-\frac{1}{a^i}+\varepsilon)=\max\{\frac{2}{a^i}-(b_i-\frac{1}{a^i}+\varepsilon), \frac{1}{a^i}\}$ and now we can apply Lemma~\ref{lemma_2} and have that the first player will get the second job in $B^{E_1}_i$ and we are left with showing that in this case the first player also gets the first job in $B^{E_1}_i$. Assume to the contrary
that the first player gets the second job but not the first job in $B^{E_1}_i$. Reduce the first player's cost for the second job in $B^{E_1}_i$ to 0 and by Lemma~\ref{lemma1}, the first player still does not get the first job in $B^{E_1}_i$. The only active players for the first job in $B^{E_1}_i$ are players $1, 2i$ and $2i+1$. Assume that the $2i$'th player gets this job (the analysis is similar if the  $(2i+1)$'th player gets it). We can now increase the $2i$'th player's dummy job's cost to $\frac{1}{a^i}$, and by Lemma~\ref{lemma3} this player will get both jobs (this case is depicted below). Then, $M$'s makespan is at least $\frac{1}{a^{i-1}}+\frac{1}{a^i}$ whereas the optimal makespan is $\frac{1}{a^i}$ which yields an approximation ratio of at least $1+a$ for the makespan.

\begin{center}
\begin{tikzpicture}
\matrix [matrix of math nodes,left delimiter=(,right delimiter=)] (m)
{
0& \dots &0   &\frac{1}{a^i}^{\textcolor{red}{@}}      &0^{\textcolor{red}{@}}  &  \frac{2}{a^i} &b_{i+1} &\frac{2}{a^{i+1}} &\frac{2}{a^{i+1}} &\dots & b_r & \frac{2}{a^r} &\frac{2}{a^r}& \dots & \dots & \dots & \dots \\
\vdots   &  \vdots  &  \vdots &\vdots   &  \vdots  &  \vdots &\vdots&  \vdots  &  \vdots &\vdots   &  \vdots  &  \vdots &\vdots &\mathcal{C}_r & \dots &\vdots &\dots\\
\dots &\dots  &\dots  &\frac{1}{a^{i-1}}^{\textcolor{blue}{*}} &  \frac{1}{a^{i-1}}   &\infty &\dots &\dots&\dots &\dots&\dots &\dots&\dots &\dots & \dots & \frac{1}{a^i}^{\textcolor{blue}{*}{\textcolor{red}{@}}} & \dots\\
\dots & \dots &\dots  &\frac{1}{a^{i-1}}  &  \infty  &\varepsilon^{\textcolor{red}{@}}  &\dots &\dots &\dots  & \dots & \dots&\dots &\dots &\dots  & \dots & \dots&\dots\\
\dots &\dots  &\dots&\dots &\dots&\dots  &\frac{1}{a^{i}}^{\textcolor{red}{@}}   &  \varepsilon^{\textcolor{red}{@}}    &\infty  &\dots&\dots &\dots&\dots &\dots & \dots & \dots & \dots\\
\dots &\dots  &\dots&\dots &\dots&\dots  &\frac{1}{a^{i}}  &  \infty   &\varepsilon^{\textcolor{red}{@}}   &\dots&\dots &\dots&\dots &\dots & \dots & \dots & \dots\\
\vdots  &  \vdots  &\vdots  &\vdots &\vdots &\vdots & \vdots &\vdots& \vdots&\ddots& \vdots&\vdots & \vdots &\vdots& \vdots&\vdots& \vdots\\
\dots &\dots  &\dots&\dots &\dots&\dots&\dots &\dots&\dots&\dots &\frac{1}{a^{r}}^{\textcolor{red}{@}}   &  \varepsilon^{\textcolor{red}{@}}    &\infty  &\dots&\dots &\dots &\dots\\
\dots &\dots  &\dots&\dots &\dots&\dots&\dots &\dots&\dots&\dots &\frac{1}{a^{r}}  &  \infty   &\varepsilon^{\textcolor{red}{@}}   &\dots&\dots &\dots &\dots\\
\vdots  &  \vdots  &\vdots  &\vdots &\vdots &\vdots & \vdots &\vdots& \vdots&\vdots& \vdots&\vdots & \vdots &\vdots& \vdots&\vdots& \vdots\\
};
\mvline[dashed, black]{m}{14};
\mvline[dashed, black]{m}{15};
\end{tikzpicture}
\end{center}

In the other scenario,  $\frac{1}{a^i}=\max\{\frac{2}{a^i}-(b_i-\frac{1}{a^i}+\varepsilon), \frac{1}{a^i}\}$. By Lemma~\ref{lemma_2}, the first player gets at least one of the first two jobs in $B^{E_1}_i$. Recall that the proof is finished if the first player gets both jobs (this is the lemma's statement). Thus, assume that the first player only gets one job. We analyze the case that this job is the first job in $B^{E_1}_i$ (the analysis is similar otherwise).
Reduce the cost of the first player for the first job in $B^{E_1}_i$ to zero and by Lemma~\ref{lemma1}, the first player still does not get the second job in $B^{E_1}_i$. Recall that second job in $B^{E_1}_i$ has only two active players, the first player and the $2i$'th player and thus, the $2i$'th player gets the second job. Next, increase the $2i$'th player's cost for his dummy job to $\frac{1}{a^i}$ and by Lemma~\ref{lemma3}, the $2i$'th player will keep both jobs (this case is depicted below ). Then, $M$'s makespan is at least $\frac{1}{a^{i-1}}+\frac{1}{a^i}$ whereas the optimal makespan is $\frac{1}{a^i}$ which yields an approximation ratio of at least $1+a$ for the makespan.

\begin{center}
\begin{tikzpicture}
\matrix [matrix of math nodes,left delimiter=(,right delimiter=)] (m)
{
0& \dots &0   &0^{\textcolor{red}{@}}      &\frac{1}{a_i}^{\textcolor{red}{@}}  &  \frac{2}{a^i} &b_{i+1} &\frac{2}{a^{i+1}} &\frac{2}{a^{i+1}} &\dots & b_r & \frac{2}{a^r} &\frac{2}{a^r}& \dots & \dots & \dots & \dots \\
\vdots   &  \vdots  &  \vdots &\vdots   &  \vdots  &  \vdots &\vdots&  \vdots  &  \vdots &\vdots   &  \vdots  &  \vdots &\vdots &\mathcal{C}_r & \dots &\vdots &\dots\\
\dots &\dots  &\dots  &\frac{1}{a^{i-1}} &  \frac{1}{a^{i-1}}^{\textcolor{blue}{*}}   &\infty &\dots &\dots&\dots &\dots&\dots &\dots&\dots &\dots & \dots & \frac{1}{a^i}^{\textcolor{blue}{*}{\textcolor{red}{@}}} & \dots\\
\dots & \dots &\dots  &\frac{1}{a^{i-1}}  &  \infty  &\varepsilon^{\textcolor{red}{@}}  &\dots &\dots &\dots  & \dots & \dots&\dots &\dots &\dots  & \dots & \dots&\dots\\
\dots &\dots  &\dots&\dots &\dots&\dots  &\frac{1}{a^{i}}^{\textcolor{red}{@}}   &  \varepsilon^{\textcolor{red}{@}}    &\infty  &\dots&\dots &\dots&\dots &\dots & \dots & \dots & \dots\\
\dots &\dots  &\dots&\dots &\dots&\dots  &\frac{1}{a^{i}}  &  \infty   &\varepsilon^{\textcolor{red}{@}}   &\dots&\dots &\dots&\dots &\dots & \dots & \dots & \dots\\
\vdots  &  \vdots  &\vdots  &\vdots &\vdots &\vdots & \vdots &\vdots& \vdots&\ddots& \vdots&\vdots & \vdots &\vdots& \vdots&\vdots& \vdots\\
\dots &\dots  &\dots&\dots &\dots&\dots&\dots &\dots&\dots&\dots &\frac{1}{a^{r}}^{\textcolor{red}{@}}   &  \varepsilon^{\textcolor{red}{@}}    &\infty  &\dots&\dots &\dots &\dots\\
\dots &\dots  &\dots&\dots &\dots&\dots&\dots &\dots&\dots&\dots &\frac{1}{a^{r}}  &  \infty   &\varepsilon^{\textcolor{red}{@}}   &\dots&\dots &\dots &\dots\\
\vdots  &  \vdots  &\vdots  &\vdots &\vdots &\vdots & \vdots &\vdots& \vdots&\vdots& \vdots&\vdots & \vdots &\vdots& \vdots&\vdots& \vdots\\
};
\mvline[dashed, black]{m}{14};
\mvline[dashed, black]{m}{15};
\end{tikzpicture}
\end{center}
\end{proof}

\subsection{Proof of Theorem ~\ref{main_thm}}\label{proof}

We now turn to prove the equivalent formulation of Theorem ~\ref{main_thm} as stated in the beginning of Section ~\ref{main_setion}.    
For a fixed $a<2$ we fix $r\in \mathbb{N}$ and $n=3r+1$. At the end of the proof we find the exact value of $r$ for the fixed $a$ which satisfies the theorem. 
Fix some mechanism $M$ for the unrelated machine scheduling problem for $n$ machines and $7r$ jobs with approximation ratio better than $1+a$. 

The proof is divided to three parts.
In the first part, Section ~\ref{first_part}, based on the assumption that $M$ has an approximation ratio better than $1+a$, we prove that there exists an instance of the form $A^E_{n,i}$ for $0 \leq i \leq r$ for which $M$ allocates all non--trivial jobs to the first player (Proposition ~\ref{the_first_player_controls_the_makespan}).

The second part is Section ~\ref{second_part}, where we use the proposition to compute some bounds on $M$'s possible approximation ratios and then we find values $b_1,...,b_n$ for which these approximation ratios are at least $1+a$ (Lemma ~\ref{same_app_ratio}).

In the last part, section ~\ref{third_part} we prove Lemma ~\ref{bk_is_atleat} which states that $b_i$ is at least $\frac{1}{a^i}$. The section also contains an example of the specific instance $\mathcal{A}_n$ for $r=3, n=10$ (Example ~\ref{example_of_main_instance}) and a table (Table ~\ref{approximation}) of some of the specific values of $n$ and $a$ for which the lower bound holds.

\subsubsection{First Part: Proposition \ref{the_first_player_controls_the_makespan}} \label{first_part}

In this section we prove Proposition ~\ref{the_first_player_controls_the_makespan} and provide an example of the proposition ~\ref{example_of_prop}.

\begin{proposition}\label{the_first_player_controls_the_makespan}
There exists an instance of the form $A^E_{n,i}$ for some $0 \leq i \leq r$ for which $M$ must allocate all the non--trivial jobs to the first player.
\end{proposition}

Recall Claim~\ref{bcd_claim}. 
Then, this proposition essentially proves that for every mechanism with approximation ratio better than $1+a$ there exists an instance of the form $A^E_{n,i}$ for some $0 \leq i \leq r$ for which the conditions of the claim holds (it actually proves a stronger guarantee that the first player gets all non--trivial jobs instead of only the non--trivial jobs in the first part of the instance). We give a full proof for completeness. 




\begin{example}[An example of Proposition ~\ref{the_first_player_controls_the_makespan} where r=2, n=7]
\label{example_of_prop}

In the case that a mechanism (for the unrelated machine scheduling problem for $7$ machines and $14$ jobs) has a better approximation ratio than $1+a$ we will show that at least one the cases in Figure ~\ref{possible_instances_n=2} will hold (Proposition ~\ref{the_first_player_controls_the_makespan}), that is the first player will get all the non--trivial jobs in the first two parts of at least one of the instances of the forms $A^E_{2,i}$ for $0 \leq i\leq 2$, that is $A^E_{2,0}, A^{E_1}_{2,1}, A^{E_2}_{2,1}, A^{E_1}_{2,2}, A^{E_2}_{2,2}$.

\begin{figure}[H]
\[
\begin{subfigure}[h]{.5\textwidth}
\centering
\begin{adjustbox}{max width=1\textwidth}
    \begin{tikzpicture}
\matrix [matrix of math nodes,left delimiter=(,right delimiter=)] (m)
{
b_1^{\textcolor{blue}{*}}   & \frac{2}{a}  &\frac{2}{a}  & b_2^{\textcolor{blue}{*}}   & \frac{2}{a^2}  &\frac{2}{a^2} & \frac{1}{a^4}^{\textcolor{blue}{*}}  & \frac{1}{a^5}^{\textcolor{blue}{*}}   & \frac{1}{a^6}^{\textcolor{blue}{*}}  &\dots\\
1  & \varepsilon &\infty &\dots&\dots&\dots&\dots&\dots&\dots&\dots\\
1  & \infty &\varepsilon &\dots &\dots &\dots&\dots&\dots&\dots&\dots\\
\dots  & \dots &\dots &\frac{1}{a} &\varepsilon &\infty&\dots&\dots&\dots&\dots\\
\dots  & \dots &\dots &\frac{1}{a} &\infty &\varepsilon&\dots&\dots&\dots&\mathcal{D}_2\\
\dots  & \dots  &\dots&\dots&\dots&\dots&\frac{1}{a^3}&\dots&\dots&\dots\\
\dots  & \dots  &\dots&\dots&\dots&\dots&\dots&\frac{1}{a^4}&\dots&\dots\\
\dots  & \dots  &\dots&\dots&\dots&\dots&\dots&\dots&\frac{1}{a^5}&\dots\\
};
\mvline[dashed, black]{m}{7};
\mvline[dashed, black]{m}{10};
    \end{tikzpicture}
    \end{adjustbox}
\caption{An instance of the form $A^E_{2,0}$}
\end{subfigure}
\]
\[
\begin{subfigure}[h]{.5\textwidth}
\begin{adjustbox}{max width=1\textwidth}
\begin{tikzpicture}
\matrix [matrix of math nodes,left delimiter=(,right delimiter=)] (m1)
{
\frac{1}{a}^{\textcolor{blue}{*}}   & c_1^{\textcolor{blue}{*}}    &\frac{2}{a}  & b_2^{\textcolor{blue}{*}}   & \frac{2}{a^2}  &\frac{2}{a^2} & \frac{1}{a^4}^{\textcolor{blue}{*}}  & \frac{1}{a^5}^{\textcolor{blue}{*}}   & \frac{1}{a^6}^{\textcolor{blue}{*}}  &\dots\\
1  & 1 &\infty &\dots&\dots&\dots&\dots&\dots&\dots&\dots\\
1  & \infty &\varepsilon &\dots &\dots &\dots&\dots&\dots&\dots&\dots\\
\dots  & \dots &\dots &\frac{1}{a} &\varepsilon &\infty&\dots&\dots&\dots&\dots\\
\dots  & \dots &\dots &\frac{1}{a} &\infty &\varepsilon&\dots&\dots&\dots&\mathcal{D}_2\\
\dots  & \dots  &\dots&\dots&\dots&\dots&\frac{1}{a^3}&\dots&\dots&\dots\\
\dots  & \dots  &\dots&\dots&\dots&\dots&\dots&\frac{1}{a^4}&\dots&\dots\\
\dots  & \dots  &\dots&\dots&\dots&\dots&\dots&\dots&\frac{1}{a^5}&\dots\\
};
\mvline[dashed, black]{m1}{7};
\mvline[dashed, black]{m1}{10};
\end{tikzpicture}
\end{adjustbox}
\caption{An instance of the form $A^{E_1}_{2,1}$}
\end{subfigure}
\begin{subfigure}[h]{.5\textwidth}
\begin{adjustbox}{max width=1\textwidth}
\begin{tikzpicture}
\matrix [matrix of math nodes,left delimiter=(,right delimiter=)] (m1)
{
\frac{1}{a}^{\textcolor{blue}{*}}   & \frac{2}{a}   &c_1^{\textcolor{blue}{*}}  & b_2^{\textcolor{blue}{*}}   & \frac{2}{a^2}  &\frac{2}{a^2} & \frac{1}{a^4}^{\textcolor{blue}{*}}  & \frac{1}{a^5}^{\textcolor{blue}{*}}   & \frac{1}{a^6}^{\textcolor{blue}{*}}  &\dots\\
1  & \varepsilon &\infty &\dots&\dots&\dots&\dots&\dots&\dots&\dots\\
1  & \infty &1 &\dots &\dots &\dots&\dots&\dots&\dots&\dots\\
\dots  & \dots &\dots &\frac{1}{a} &\varepsilon &\infty&\dots&\dots&\dots&\dots\\
\dots  & \dots &\dots &\frac{1}{a} &\infty &\varepsilon&\dots&\dots&\dots&\mathcal{D}_2\\
\dots  & \dots  &\dots&\dots&\dots&\dots&\frac{1}{a^3}&\dots&\dots&\dots\\
\dots  & \dots  &\dots&\dots&\dots&\dots&\dots&\frac{1}{a^4}&\dots&\dots\\
\dots  & \dots  &\dots&\dots&\dots&\dots&\dots&\dots&\frac{1}{a^5}&\dots\\
};
\mvline[dashed, black]{m1}{7};
\mvline[dashed, black]{m1}{10};
\end{tikzpicture}
\end{adjustbox}
\caption{An instance of the form $A^{E_2}_{2,1}$}
\end{subfigure}
\]
\[
\begin{subfigure}[h]{.5\textwidth}
\begin{adjustbox}{max width=1\textwidth}
\begin{tikzpicture}
\matrix [matrix of math nodes,left delimiter=(,right delimiter=)] (m)
{
0     &\dots  & \frac{1}{a^2}^{\textcolor{blue}{*}}   & c_2^{\textcolor{blue}{*}}  &\frac{2}{a^2} & \frac{1}{a^4}^{\textcolor{blue}{*}}  & \frac{1}{a^5}^{\textcolor{blue}{*}}   & \frac{1}{a^6}^{\textcolor{blue}{*}}  &\dots\\
\dots   &\dots &\dots&\dots&\dots&\dots&\dots&\dots&\dots\\
\dots   &\dots &\dots &\dots &\dots&\dots&\dots&\dots&\dots\\
\dots   &\dots &\frac{1}{a} &\frac{1}{a} &\infty&\dots&\dots&\dots&\dots\\
\dots   &\dots &\frac{1}{a} &\infty &\varepsilon&\dots&\dots&\dots&\mathcal{D}_2\\
\dots   &\dots&\dots&\dots&\dots&\frac{1}{a^3}&\dots&\dots&\dots\\
\dots    &\dots&\dots&\dots&\dots&\dots&\frac{1}{a^4}&\dots&\dots\\
\dots   &\dots&\dots&\dots&\dots&\dots&\dots&\frac{1}{a^5}&\dots\\
};
\mvline[dashed, black]{m}{6};
\mvline[dashed, black]{m}{9};
\end{tikzpicture}
\end{adjustbox}
\caption{An instance of the form $A^{E_1}_{2,2}$}
\end{subfigure}
\begin{subfigure}[h]{.5\textwidth}
\begin{adjustbox}{max width=1\textwidth}
\begin{tikzpicture}
\matrix [matrix of math nodes,left delimiter=(,right delimiter=)] (m)
{
0     &\dots  & \frac{1}{a^2}^{\textcolor{blue}{*}}   &\frac{2}{a^2} &c_2^{\textcolor{blue}{*}}  & \frac{1}{a^4}^{\textcolor{blue}{*}}  & \frac{1}{a^5}^{\textcolor{blue}{*}}   & \frac{1}{a^6}^{\textcolor{blue}{*}}  &\dots\\
\dots   &\dots &\dots&\dots&\dots&\dots&\dots&\dots&\dots\\
\dots   &\dots &\dots &\dots &\dots&\dots&\dots&\dots&\dots\\
\dots   &\dots &\frac{1}{a} &\varepsilon &\infty&\dots&\dots&\dots&\dots\\
\dots   &\dots &\frac{1}{a} &\infty &\frac{1}{a}&\dots&\dots&\dots&\mathcal{D}_2\\
\dots   &\dots&\dots&\dots&\dots&\frac{1}{a^3}&\dots&\dots&\dots\\
\dots    &\dots&\dots&\dots&\dots&\dots&\frac{1}{a^4}&\dots&\dots\\
\dots   &\dots&\dots&\dots&\dots&\dots&\dots&\frac{1}{a^5}&\dots\\
};
\mvline[dashed, black]{m}{6};
\mvline[dashed, black]{m}{9};
\end{tikzpicture}
\end{adjustbox}
\caption{An instance of the form $A^{E_2}_{2,2}$}
\end{subfigure}
\]
\caption{ These are all the possible cases described in Example ~\ref{example_of_prop}. In case that $r=2, n=7$, a mechanism with better approximation ratio than $1+a$ must allocate all the non--trivial jobs to the first player in at least one of these instances. Where $c_1 = \max\{\frac{2}{a}-(b_1-\frac{1}{a}+\varepsilon), \frac{1}{a}\} $ and $c_2=\max\{\frac{2}{a^2}-(b_2-\frac{1}{a^2}+\varepsilon)\, \frac{1}{a^2}\} $.}
\label{possible_instances_n=2}
\end{figure}
\end{example}

\begin{proof}[Proof of Proposition \ref{the_first_player_controls_the_makespan}]
In order to prove the proposition we first state and prove two lemmas (Lemma ~\ref{all_jobs_in_C} and Lemma ~\ref{semi_dummy_goes_to_epsilon}).
\begin{lemma}\label{all_jobs_in_C}
Consider an instance of the form $A^E_{n,i}$ for $0 \leq i \leq r$ and let $M$ be a mechanism with approximation ratio better than $1+a$ that allocates all the non--trivial jobs in the first part of the instance to the first player. Then, $M$ must also allocate all jobs in $C$ (all jobs in the second part of the instance) to the first player.
\end{lemma}

\begin{proof}[Proof of Lemma \ref{all_jobs_in_C}]
Conversely, suppose that $M$ does allocate to the first player all the non--trivial jobs in the first part of the instance of the form $A^E_{n,i}$ but does not allocate to him all the jobs in $C$. Denote by $1 \leq j\leq r$ the first job in $C$ that is not allocated to the first player by $M$. Depicted below is this case for $i\geq 1$.

\begin{center}
\begin{tikzpicture}
\matrix [matrix of math nodes,left delimiter=(,right delimiter=)] (m)
{
0  &  \dots  &  \frac{1}{a^k}^{\textcolor{blue}{*}}   &   {\max\{\frac{2}{a^i}-(b_i-\frac{1}{a^i}+\varepsilon), \frac{1}{a^i}\}}^{\textcolor{blue}{*}}  &  \frac{2}{a^k}  &  b_{k+1}^{\textcolor{blue}{*}}  &  \dots  &  b_r^{\textcolor{blue}{*}}  &  \frac{1}{a^{r+1}}^{\textcolor{blue}{*}}  &   \dots  & \frac{1}{a^{r+j}} &  \dots  &  \frac{1}{a^{2r}}  &  \dots\\
\vdots  &   \vdots  &  \vdots  &  \vdots  &   \vdots  &  \vdots  &  \vdots  &  \vdots  &   \vdots  &  \dots  &  \vdots  &  \dots &  \vdots  &  \dots\\
\dots  &  \dots  &   \frac{1}{a^{k-1}}   &  \frac{1}{a^{k-1}}  &  \infty  &   \dots  &  \dots  &  \dots  &  \dots  &   \dots  &  \dots  &  \dots&  \dots  &  \dots\\
\dots  &  \dots  &   \frac{1}{a^{k-1}}   &  \infty  &  \varepsilon  &   \dots  &  \dots  &  \dots  &  \dots  &   \dots  &  \dots  &  \dots&  \dots  &  \dots\\
\dots  &  \dots  &   \dots   &  \dots  &  \dots  &   \frac{1}{a^k}  &  \dots  &  \dots  &  \dots  &   \dots  &  \dots &  \dots  &  \dots  &  \mathcal{D}_r\\
\dots  &  \dots  &   \dots   &  \dots  &  \dots  &   \dots  &  \ddots  &  \dots  &  \dots  &   \dots  &  \dots  &  \dots&  \dots  &  \dots\\
\dots  &  \dots  &   \dots   &  \dots  &  \dots  &   \dots  &  \dots  &  \frac{1}{a^{r-1}}  &  \dots  &   \dots  &  \dots  &  \dots&  \dots  &  \dots\\
\dots  &  \dots  &   \dots   &  \dots  &  \dots  &   \dots  &  \dots  &  \dots  &  \frac{1}{a^{r}}  &   \dots  &  \dots  &  \dots&  \dots  &  \dots\\
\vdots  &   \vdots  &  \vdots  &  \vdots  &   \vdots  &  \vdots  &  \vdots  &  \vdots  &   \dots  &  \ddots  &  \dots  &  \dots&  \dots  &  \dots\\
\dots  &  \dots  &   \dots   &  \dots  &  \dots  &   \dots  &  \dots  &  \dots  &  \dots &   \dots  &  \frac{1}{a^{r+j-1}}^{\textcolor{blue}{*}}    &  \dots&  \dots  &  \dots\\
\vdots  &   \vdots  &  \vdots  &  \vdots  &   \vdots  &  \vdots  &  \vdots  &  \vdots  &   \dots  &  \dots  &  \dots  &  \ddots&  \dots  &  \dots\\
\dots  &  \dots  &   \dots   &  \dots  &  \dots  &   \dots  &  \dots  &  \dots  &  \dots &   \dots  &  \dots   &  \dots&  \frac{1}{a^{2r-1}}  &  \dots\\
};
\mvline[dashed, black]{m}{9};
\mvline[dashed, black]{m}{14};
\end{tikzpicture}
\end{center}

Now, reduce the first player's cost for all the non--trivial jobs in the first part of the instance and for the first $j-1$ jobs in $C$ to 0. By Lemma~\ref{lemma1}, the first player still does not get the $j$'th job in $C$ and since $M$ has a finite approximation ratio it will allocate it to the $(2r+j+1)$'th player (the first player and the $(2r+j+1)$'th player are the two only active players for the $j$'th job in $C$). Next, we increase the $(2r+j+1)$'th player's dummy job's to $\frac{1}{a^{r+j}}$ and by Lemma~\ref{lemma3} this player's allocation remains the same. $M$'s makespan is at least $\frac{1}{a^{r+j-1}}+\frac{1}{a^{r+j}}$ whereas the optimal makespan is $1+a$ (this is depicted below)
which results in $M$ having a makespan of $1+a$.

\begin{center}
\begin{tikzpicture}
\matrix [matrix of math nodes,left delimiter=(,right delimiter=)] (m)
{
0^{\textcolor{red}{@}}  &  \dots  &  0^{\textcolor{red}{@}}   &   0^{\textcolor{red}{@}}  &  \dots  &  \frac{1}{a^{r+j}}^{\textcolor{red}{@}}  &  \frac{1}{a^{r+j+1}}  &  \dots  &   \frac{1}{a^{2r}}  &   \dots  & \dots &  \dots\\
\vdots  &   \vdots  &  \vdots  &  \vdots  &   \vdots  &  \vdots  &  \vdots  &  \vdots  &   \vdots  &  \dots  &  \dots  &  \dots \\
\dots  &  \dots  &   \dots   &  \frac{1}{a^r}   &  \dots  &  \dots  &   \dots  &  \dots  &  \dots  &  \dots &   \dots  &  \dots\\
\vdots  &  \vdots  &   \vdots   &  \vdots  &  \ddots  &   \vdots  &  \vdots  &  \vdots  &  \vdots  &   \vdots  &  \vdots&  \vdots\\
\dots  &  \dots  &   \dots   &  \dots  &  \dots  &   \frac{1}{a^{r+j-1}}^{\textcolor{blue}{*}}  &  \dots  &  \dots  &  \dots  &   \dots  &  \frac{1}{a^{r+j}}^{\textcolor{blue}{*}{\textcolor{red}{@}}}  &  \dots\\
\dots  &  \dots  &   \dots   &  \dots  &  \dots  &   \dots  &  \frac{1}{a^{r+j}}^{\textcolor{red}{@}}  &  \dots  &  \dots  &   \dots  &  \dots  &  \dots\\
\vdots  &   \vdots  &  \vdots  &  \vdots  &   \vdots  &  \vdots  &  \vdots  &  \ddots  &   \dots  &  \dots  &  \dots  &  \dots\\
\dots  &  \dots  &   \dots   &  \dots  &  \dots  &   \dots  &  \dots  &  \dots  &  \frac{1}{a^{2r-1}}^{\textcolor{red}{@}}  &   \dots  &  \dots  &  \dots\\
};
\mvline[dashed, black]{m}{4};
\mvline[dashed, black]{m}{10};
\end{tikzpicture}
\end{center}
\end{proof}

Lemma ~\ref{all_jobs_in_C} is a customized version of Lemma~\ref{c_lemma}. The remaining lemma handles the case that the first job in a block is allocated to the $2i_1$'th player, it is similar to the case that the first job is instead allocated to the $2i_1+1$'th player, the needed changes are written in parentheses.   

\begin{lemma}\label{semi_dummy_goes_to_epsilon}
Let $A'_{n,i}$ be an instance of the form $\mathcal{A}_{n,i}$ and suppose that $M$ allocates the first job in the $i_1$'th block of $A'_{n,i}$ to the $2i_1$'th ($2i_1+1$'th) player. Then, $M$ also allocates the second (third) job in the $i_1$'th block to the $2i_1$'th ($2i_1+1$'th) player.
\end{lemma}

\begin{proof}[Proof of Lemma \ref{semi_dummy_goes_to_epsilon}]
Suppose, contrary to our claim, that $M$ allocates the first job in $B_{i_1}$ to the $2i_1$'th player but it does not  allocate him the second job in $B_{i_1}$. Recall that the second job in $B_{i_1}$ has exactly two active players, the first player and the $2i_1$'th player. Thus, since $M$ has a finite approximation ratio it must allocate this job to the first player. 

Now, reduce the $2i_1$'th player's cost for the first job in $B_{i_1}$ to 0. By Lemma~\ref{lemma1}, the $2i_1$'th player still does not get the second job in $B_{i_1}$ and using the same argument as before (since there are only two active players for this job) the first player will get it.
We can now increase the first player's dummy job's cost to $\frac{1}{a^{i_1}}$ and by Lemma~\ref{lemma3} the first player will keep both his dummy job and the second job in $B_{i_1}$ which results in a makespan of $\frac{3}{a^{i_1}}$ whereas the optimal makespan is $\frac{1}{a^{i_1}}$, as depicted below. Thus, $M$ has an approximation ratio of 3 (which is worse than $1+a$).

\begin{center}
\begin{tikzpicture}
\matrix [matrix of math nodes,left delimiter=(,right delimiter=)] (m)
{
\dots  &  0^{\textcolor{red}{@}}   &  \frac{2}{a^{i_{1}-1}}  &  \frac{2}{a^{i_{1}-1}}  &  b_{i_1}  &   \frac{2}{a^{i_1}}^{\textcolor{blue}{*}}  &  \frac{2}{a^{i_1}}  &  b_{{i_1}+1}  &  \dots  &   \dots  &  \frac{1}{a^{i_1}}^{\textcolor{blue}{*}{\textcolor{red}{@}} }  &  \infty&  \dots  &  \dots  &  \dots  &   \dots\\
\vdots  &   \vdots  &  \vdots  &  \vdots  &   \vdots  &  \vdots  &  \vdots  &  \vdots  &   \vdots  &  \dots  &  \vdots  &  \ddots  &   \ddots  &  \ddots  &   \ddots  &  \vdots  \\
\dots  &  \frac{1}{a^{i_{1}-2}}  &   \varepsilon^{\textcolor{red}{@}}   &  \infty  &  \dots  &   \dots  &  \dots  &  \dots  &  \dots  &   \dots  &  \infty  &  \ddots  &  \ddots  &   \ddots  &   \ddots &  \vdots\\
\dots  &  \frac{1}{a^{i_{1}-2}}  &   \infty   &   \varepsilon^{\textcolor{red}{@}}   &  \dots  &   \dots  &  \dots  &  \dots  &  \dots  &   \dots  &  \vdots  &  \ddots  &  \ddots  &   \ddots  &   \ddots &  \vdots\\
\dots &   \dots  &  \dots  &  \dots  &   0^{\textcolor{red}{@}}   &  \varepsilon^{\textcolor{red}{@}}   &  \infty  &  \dots  &   \dots  &  \mathcal{C}_r  &   \vdots  &  \ddots  &   \ddots  &  \ddots  &   \ddots  &  \vdots\\
\dots &   \dots  &  \dots  &  \dots  &   \frac{1}{a^{i_{1}-1}}  &  \infty  &  \varepsilon^{\textcolor{red}{@}}   &  \dots  &   \dots  &  \dots  &   \vdots  &  \ddots  &   \ddots  &  \ddots  &   \ddots  &  \vdots\\
\dots  &   \dots  &  \dots  &  \dots  &   \dots  &  \dots  &  \dots  &  \frac{1}{a^{i_1}}^{\textcolor{red}{@}}    &   \dots  &  \dots  &  \vdots  &  \ddots  &   \ddots  &  \ddots  &   \ddots  &  \vdots\\
\dots  &   \dots  &  \dots  &  \dots  &   \dots  &  \dots  &  \dots  &  \frac{1}{a^{i_1}}   &   \dots  &  \dots  &  \vdots  &  \ddots  &   \ddots  &  \ddots  &   \ddots  &  \vdots\\
\vdots  &   \vdots  &  \vdots  &  \vdots  &   \vdots  &  \vdots  &  \vdots  &  \vdots  &   \vdots  &  \dots  &  \vdots  &  \ddots  &   \ddots  &  \ddots  &   \ddots  &  \vdots  \\
};
\mvline[dashed, black]{m}{10};
\mvline[dashed, black]{m}{11};
\end{tikzpicture}
\end{center}
\end{proof}

Now, we can start the proof of Proposition ~\ref{the_first_player_controls_the_makespan}. Consider $M$'s possible allocations for the instance $\mathcal{A}_n$. If $M$ allocates all jobs in $B$ (all the non--trivial jobs in the first part of $\mathcal{A}_n$) to the first player, then, by Lemma ~\ref{all_jobs_in_C}, we have that $M$ also allocates all the jobs in $C$ to the first player. I.e., we have an instance of the form $A^E_{n,0}$ in which $M$ allocates all non--trivial jobs to the first player and the proposition follows (for an example of this case, see Figure ~\ref{example_of_prop} case $(a)$).

In the other case, $M$ does not allocate all the jobs in $B$ to the first player. Denote by $1 \leq i_1\leq r$ the index of the first block in $\mathcal{B}_r$ whose first job (the $3\cdot(i_1-1)+1$ job) is not allocated to the first player. By reducing the cost of the first player for all the non--trivial jobs he got in the first $i_1-1$ blocks we obtain an instance of the form $\mathcal{A}_{n,i_1}$ in which by Lemma~\ref{lemma1}, the first player does not get the first job in $B_{i_1}$.   Recall that the only active players for the first job in $B_{i_1}$ are players numbered $1, 2i_1$ and $(2i_1+1)$ and since the first player is not allocated this job one of the other two is. Assume that $M$ allocates this job to the $2i_1$'th player (the analysis is similar if instead $M$ allocates this job to the $2i_1 +1 $'th player). Then by Lemma ~\ref{semi_dummy_goes_to_epsilon}, it follows that the $2i_1$'th player also gets the second job in $B_{i_1}$. Now, we have an instance of the form $\mathcal{A}_{n,i_1}$ for which $M$ allocates the first two jobs in $B_{i_1}$ to the $2i_1$'th player as depicted below. 

\begin{center}
\[
\begin{tikzpicture}[baseline=(m-3-1.base)]
\matrix [matrix of math nodes,left delimiter=(,right delimiter=)] (m)
{
0^{\textcolor{blue}{*}}  &   \frac{2}{a}  &  \frac{2}{a}  &  \dots  &   0{\textcolor{blue}{*}}  &  \frac{2}{a^{i_{1}-1}}  &  \frac{2}{a^{i_{1}-1}}  &  b_{i_1}  &   \frac{2}{a^{i_1}}  &  \frac{2}{a^{i_1}}  &  \dots  &  b_r  &   \frac{2}{a^r}  &  \frac{2}{a^r}\\
1  &   \varepsilon  &  \infty  &  \dots  &   \dots  &  \dots  &  \dots  &  \dots  &   \dots  &  \dots  &  \dots  &  \dots  &   \dots  &  \dots\\
1  &   \infty  &  \varepsilon  &  \dots  &   \dots  &  \dots  &  \dots  &  \dots  &   \dots  &  \dots  &  \dots  &  \dots  &   \dots  &  \dots \\
\vdots  &   \vdots  &  \vdots  &  \ddots  &   \vdots  &  \vdots  &  \vdots  &  \vdots  &   \vdots  &  \vdots  &  \vdots  &   \vdots  &  \vdots  &  \vdots\\
\infty  &   \dots  &  \dots  &  \dots  &   \frac{1}{a^{i_{1}-2}}  &  \varepsilon  &  \infty  &  \dots  &   \dots  &  \dots  &  \dots  &  \dots  &   \dots  &  \dots  \\
\infty  &   \dots  &  \dots  &  \dots  &   \frac{1}{a^{i_{1}-2}}  &  \infty &  \varepsilon   &  \dots  &   \dots  &  \dots  &  \dots  &  \dots  &   \dots  &  \dots  \\
\infty  &   \dots  &  \dots  &  \dots  &   \dots  &  \dots  &  \dots  &  \frac{1}{a^{i-1}}^{\textcolor{blue}{*}}   &   \varepsilon^{\textcolor{blue}{*}}  &  \infty  &  \dots  &  \dots  &   \dots  &  \dots\\
\infty  &   \dots  &  \dots  &  \dots  &   \dots  &  \dots  &  \dots  &  \frac{1}{a^{i-1}}  &   \infty  &  \varepsilon  &  \dots  &  \dots  &   \dots  &  \dots  \\
\vdots  &   \vdots  &  \vdots  &  \vdots  &   \vdots  &  \vdots  &  \vdots  &  \vdots  &   \vdots  &  \vdots  &  \ddots  &  \vdots  &   \vdots  &  \vdots  \\
\infty  &   \dots  &  \dots  &  \dots  &   \dots  &  \dots  &  \dots  &  \dots  &   \dots  &  \dots  &  \dots  &  \frac{1}{a^{r-1}}  &   \varepsilon  &  \infty  \\
\infty  &   \dots  &  \dots  &  \dots  &   \dots  &  \dots  &  \dots  &  \dots  &   \dots  &  \dots  &  \dots  &  \frac{1}{a^{r-1}}  &   \infty  &  \varepsilon \\
\vdots  &   \vdots  &  \vdots  &  \vdots  &   \vdots  &  \vdots  &  \vdots  &  \vdots  &   \vdots  &  \vdots  &  \vdots  &  \vdots  &   \vdots  &  \vdots  \\
};
\end{tikzpicture}
\]
\end{center}

Next, we apply transition $E_1$ on $B_{i_1}$ (in the case that the $(2{i_1}+1)$'th player gets the first job in $B_{i_1}$ we will apply transition $E_2$ on $B_{i_1}$) and get an instance of the form $A^{E_1}_{n,i_1}$. By Lemma ~\ref{transition_lemma_1}, $M$ allocates the first two jobs in $B^{E_1}_{i_1}$ to the first player. There are two possible cases regarding $M$'s allocation of the remaining non--trivial jobs in the first part of the instance. In the first case, $M$ allocates all these jobs to the first player, or we have that $i_1=r$ and there are no such jobs; either way, by applying Lemma ~\ref{all_jobs_in_C} we have found an instance of the form $A^{E_1}_{n,i_1}$ for which $M$ allocates all the non--trivial jobs to the first player (this is depicted below) and the proof of the proposition is complete.

\begin{center}
\begin{tikzpicture}
\matrix [matrix of math nodes,left delimiter=(,right delimiter=)] (m)
{
0  &  \dots  &  \frac{1}{a^{i_1}}^{\textcolor{blue}{*}}   &   {\max\{\frac{2}{a^{i_1}}-(b_{i_1}-\frac{1}{a^{i_1}}+\varepsilon), \frac{1}{a^{i_1}}\}}^{\textcolor{blue}{*}} -\varepsilon)^{\textcolor{blue}{*}}  &  \frac{2}{a^{i_1}}  &  b_{{i_1}+1}^{\textcolor{blue}{*}}  &  \dots  &  b_r^{\textcolor{blue}{*}}  &  \frac{1}{a^{r+1}}^{\textcolor{blue}{*}}   &   \dots  &  \frac{1}{a^{2r}}^{\textcolor{blue}{*}}   &  \dots\\
\vdots  &   \vdots  &  \vdots  &  \vdots  &   \vdots  &  \vdots  &  \vdots  &  \vdots  &   \vdots  &  \dots  &  \vdots  &  \dots\\
\dots  &  \dots  &   \frac{1}{a^{{i_1}-1}}   &  \frac{1}{a^{{i_1}-1}}  &  \infty  &   \dots  &  \dots  &  \dots  &  \dots  &   \dots  &  \dots  &  \dots\\
\dots  &  \dots  &   \frac{1}{a^{{i_1}-1}}   &  \infty  &  \varepsilon  &   \dots  &  \dots  &  \dots  &  \dots  &   \dots  &  \dots  &  \dots\\
\dots  &  \dots  &   \dots   &  \dots  &  \dots  &   \frac{1}{a^{i_1}}  &  \dots  &  \dots  &  \dots  &   \dots  &  \dots  &  \mathcal{D}_r\\
\dots  &  \dots  &   \dots   &  \dots  &  \dots  &   \dots  &  \ddots  &  \dots  &  \dots  &   \dots  &  \dots  &  \dots\\
\dots  &  \dots  &   \dots   &  \dots  &  \dots  &   \dots  &  \dots  &  \frac{1}{a^{r-1}}  &  \dots  &   \dots  &  \dots  &  \dots\\
\dots  &  \dots  &   \dots   &  \dots  &  \dots  &   \dots  &  \dots  &  \dots  &  \frac{1}{a^{r}}  &   \dots  &  \dots  &  \dots\\
\vdots  &   \vdots  &  \vdots  &  \vdots  &   \vdots  &  \vdots  &  \vdots  &  \vdots  &   \dots  &  \ddots  &  \dots  &  \dots\\
\dots  &  \dots  &   \dots   &  \dots  &  \dots  &   \dots  &  \dots  &  \dots  &  \dots &   \dots  &  \frac{1}{a^{2r-1}}   &  \dots\\
};
\mvline[dashed, black]{m}{9};
\mvline[dashed, black]{m}{12};
\end{tikzpicture}
\end{center}

In the other case, $M$ allocates the first two jobs in $B^{E_1}_{i_1}$ to the first player but does not allocate all the remaining non--trivial jobs to the first player. Denote by $i_1 < i_2\leq r$ the index of the first block among the blocks with a non--trivial job {$B_{{i_1}+1},...,B_r$} whose first job (the $3\cdot(i_1-1)+1$ job) is not allocated  to the first player.

As before, we decrease the first player's cost for all the non--trivial jobs he got in the first $i_2-1$ blocks to obtain an instance of the form $\mathcal{A}_{n,i_2}$ in which by Lemma~\ref{lemma1}, the first player does not get the first job in $B_{i_2}$. 
Suppose that the the $2i_2$'th player get this job, then again we apply transition $E_1$ on $B_{i_2}$  and by Lemma ~\ref{transition_lemma_1} we have an instance of the form $A^{E_1}_{n,i_2}$ for which $M$ allocates the first two jobs in $B^{E_1}_{i_2}$ to the first player. In the case that the first player also gets all the non--trivial jobs in the first part of the instance (or there are no such jobs, since $i_2=r$) we are done (more accurately, we apply Lemma ~\ref{all_jobs_in_C} and then we are done), this case is depicted below.

\begin{center}
\begin{tikzpicture}
\matrix [matrix of math nodes,left delimiter=(,right delimiter=)] (m)
{
0  &  \dots  &  \frac{1}{a^{i_2}}^{\textcolor{blue}{*}}   &   {\max\{\frac{2}{a^{i_2}}-(b_{i_2}-\frac{1}{a^{i_2}}+\varepsilon), \frac{1}{a^{i_2}}\}}^{\textcolor{blue}{*}} -\varepsilon)^{\textcolor{blue}{*}}  &  \frac{2}{a^{i_2}}  &  b_{{i_2}+1}^{\textcolor{blue}{*}}  &  \dots  &  b_r^{\textcolor{blue}{*}}  &  \frac{1}{a^{r+1}}^{\textcolor{blue}{*}}   &   \dots  &  \frac{1}{a^{2r}}^{\textcolor{blue}{*}}   &  \dots\\
\vdots  &   \vdots  &  \vdots  &  \vdots  &   \vdots  &  \vdots  &  \vdots  &  \vdots  &   \vdots  &  \dots  &  \vdots  &  \dots\\
\dots  &  \dots  &   \frac{1}{a^{{i_2}-1}}   &  \frac{1}{a^{{i_2}-1}}  &  \infty  &   \dots  &  \dots  &  \dots  &  \dots  &   \dots  &  \dots  &  \dots\\
\dots  &  \dots  &   \frac{1}{a^{{i_2}-1}}   &  \infty  &  \varepsilon  &   \dots  &  \dots  &  \dots  &  \dots  &   \dots  &  \dots  &  \dots\\
\dots  &  \dots  &   \dots   &  \dots  &  \dots  &   \frac{1}{a^{i_2}}  &  \dots  &  \dots  &  \dots  &   \dots  &  \dots  &  \mathcal{D}_r\\
\dots  &  \dots  &   \dots   &  \dots  &  \dots  &   \dots  &  \ddots  &  \dots  &  \dots  &   \dots  &  \dots  &  \dots\\
\dots  &  \dots  &   \dots   &  \dots  &  \dots  &   \dots  &  \dots  &  \frac{1}{a^{r-1}}  &  \dots  &   \dots  &  \dots  &  \dots\\
\dots  &  \dots  &   \dots   &  \dots  &  \dots  &   \dots  &  \dots  &  \dots  &  \frac{1}{a^{r}}  &   \dots  &  \dots  &  \dots\\
\vdots  &   \vdots  &  \vdots  &  \vdots  &   \vdots  &  \vdots  &  \vdots  &  \vdots  &   \dots  &  \ddots  &  \dots  &  \dots\\
\dots  &  \dots  &   \dots   &  \dots  &  \dots  &   \dots  &  \dots  &  \dots  &  \dots &   \dots  &  \frac{1}{a^{2r-1}}   &  \dots\\
};
\mvline[dashed, black]{m}{9};
\mvline[dashed, black]{m}{12};
\end{tikzpicture}
\end{center}

In the other case, there exists a non--trivial job in the first part that is not allocated to the first player. We continue in the same manner as before: defining $i_2<i_3 \leq r$, obtaining an instance of the form $\mathcal{A}_{n,i_3}$ in which the first player does not get the first job in $B_{i_3}$. Then, we apply transition $E_1$ (or $E_2$) and obtain an instance of the form $A^{E_1}_{n,i_3}$ (or $A^{E_2}_{n,i_3}$) in which the first player does get the first two jobs in $B^{E_1}_{i_3}$. Again there are two possible cases, in the first we are done and in the second we continue to repeat the same process for $i_4$ and so on.

Observe that for every $1 \leq j\leq r-1$ it holds that $i_j<i_{j+1}$ and that the process is stopped if $i_j=r$. Thus, this process will stop after at most $r$ times. When the process stops we have completed the proof of the proposition.
\end{proof}

\subsubsection{Second Part: Bounding $M$'s Approximation Ratio}\label{second_part}

Next, we use Proposition ~\ref{the_first_player_controls_the_makespan} to compute bounds on $M$'s possible approximation ratio and in Lemma ~\ref{same_app_ratio} we prove that these bounds are at least $1+a$. 

Consider an instance of the form $A^E_{n,k}$ for $ \leq k\leq r$ for which $M$ allocates all the non--trivial jobs to the first player. We achieve a lower bound on $M$'s approximation ratio by increasing the first player's cost for his dummy job to the optimal makespan. By Lemma~\ref{lemma3} we have that the first player keeps all his jobs which results in an approximation ratio of \eqref{eq1} for the case that  $1 \leq k \leq r$ and in an approximation ratio of \eqref{equ2} for the case that $k=0$. The former case is depicted in Figure ~\ref{first_app} and the latter is depicted in Figure ~\ref{second_app}.  

\begin{equation} \label{eq1}
   \frac{\frac{1}{a^{k-1}}+\frac{1}{a^k}+{\max\{\frac{2}{a^{k}}-(b_i-\frac{1}{a^{k}}+\varepsilon), \frac{1}{a^{k}}\}}+b_{k+1}+...+b_r+\frac{1}{a^{r+1}}+...+\frac{1}{a^{2r}}}{\frac{1}{a^{k-1}}} \\
\end{equation}

\begin{equation} \label{equ2}
1+b_1+...+b_r+\frac{1}{a^{r+1}}+...+\frac{1}{a^{2r}}
\end{equation}

\begin{figure}[H]
\centering
\begin{tikzpicture}
\matrix [matrix of math nodes,left delimiter=(,right delimiter=)] (m)
{
0^{\textcolor{red}{@}}  &  \dots  &  \frac{1}{a^k}^{\textcolor{blue}{*}}   &   {\max\{\frac{2}{a^{k}}-(b_k-\frac{1}{a^{k}}+\varepsilon), \frac{1}{a^{k}}\}}^{\textcolor{blue}{*}}  &  \frac{2}{a^k}  &  b_{k+1}^{\textcolor{blue}{*}}  &  \dots  &  b_r^{\textcolor{blue}{*}}  &  \frac{1}{a^{r+1}}^{\textcolor{blue}{*}}  &   \dots  &  \frac{1}{a^{2r}}^{\textcolor{blue}{*}}  &\frac{1}{a^{k-1}}^{\textcolor{blue}{*}{\textcolor{red}{@}}} &  \dots\\
\vdots  &   \vdots  &  \vdots  &  \vdots  &   \vdots  &  \vdots  &  \vdots  &  \vdots  &   \vdots  &  \dots  &  \vdots  &  \dots&  \dots\\
\dots  &  \dots  &   \frac{1}{a^{k-1}}   &  \frac{1}{a^{k-1}}^{\textcolor{red}{@}}  &  \infty  &   \dots  &  \dots  &  \dots  &  \dots  &   \dots  &  \dots  &  \dots&  \dots\\
\dots  &  \dots  &   \frac{1}{a^{k-1}}^{\textcolor{red}{@}}   &  \infty  &  \varepsilon^{\textcolor{red}{@}}  &   \dots  &  \dots  &  \dots  &  \dots  &   \dots  &  \dots  &  \dots&  \dots\\
\dots  &  \dots  &   \dots   &  \dots  &  \dots  &   \frac{1}{a^k}^{\textcolor{red}{@}}  &  \dots  &  \dots  &  \dots  &   \dots  &  \dots  &  \dots&  \dots\\
\dots  &  \dots  &   \dots   &  \dots  &  \dots  &   \dots  &  \ddots  &  \dots  &  \dots  &   \dots  &  \dots  &  \dots&  \dots\\
\dots  &  \dots  &   \dots   &  \dots  &  \dots  &   \dots  &  \dots  &  \frac{1}{a^{r-1}}^{\textcolor{red}{@}}  &  \dots  &   \dots  &  \dots  &  \dots&  \dots\\
\dots  &  \dots  &   \dots   &  \dots  &  \dots  &   \dots  &  \dots  &  \dots  &  \frac{1}{a^{r}}^{\textcolor{red}{@}}  &   \dots  &  \dots  &  \dots&  \dots\\
\vdots  &   \vdots  &  \vdots  &  \vdots  &   \vdots  &  \vdots  &  \vdots  &  \vdots  &   \dots  &  \ddots  &  \dots  &  \dots&  \dots\\
\dots  &  \dots  &   \dots   &  \dots  &  \dots  &   \dots  &  \dots  &  \dots  &  \dots &   \dots  &  \frac{1}{a^{2r-1}}^{\textcolor{red}{@}}   &  \dots&  \dots\\
};
\mvline[dashed, black]{m}{9};
\mvline[dashed, black]{m}{12};
\end{tikzpicture}
\caption{In this case we achieve a lower bound on $M$'s approximation ratio when $1 \leq k \leq r$.}
\label{first_app}
\end{figure}

\begin{figure}[H]
\centering
\begin{tikzpicture}
\matrix [matrix of math nodes,left delimiter=(,right delimiter=)] (m)
{
b_1^{\textcolor{blue}{*}}  &   \frac{2}{a}  &  \frac{2}{a}  &  \dots  &   b_r^{\textcolor{blue}{*}}  &  \frac{2}{a^r}  &  \frac{2}{a^r}  &\frac{1}{a^{r+1}}  &  \dots  &  \frac{1}{a^{2r}}  &  1^{\textcolor{blue}{*}\textcolor{red}{@}}  &  \infty  &  \dots  &  \dots  &  \dots  &  \dots  &  \dots  &  \dots  &  \infty\\
1^{\textcolor{red}{@}}  &   \varepsilon^{\textcolor{red}{@}}  &  \infty  &  \dots  &  \dots  &   \dots  &  \dots  &  \dots  &  \dots  &   \dots  &  \infty  &  \ddots  &  \ddots  &   \ddots  &  \ddots  &  \ddots  &   \ddots  &  \ddots  &  \vdots\\
1  &   \infty  &  \varepsilon^{\textcolor{red}{@}}  &  \dots  &   \dots  &  \dots  &  \dots  &  \dots  &   \dots  &  \dots  &  \vdots  &  \ddots  &   \ddots  &  \ddots  &  \ddots  &   \ddots  &  \ddots  &  \ddots  &  \vdots  \\
\vdots  &   \vdots  &  \vdots  &  \ddots  &   \vdots  &  \vdots  &  \vdots  &  \vdots  &   \vdots  &  \vdots  &  \vdots  &  \ddots  &   \ddots  &  \ddots  &  \ddots  &   \ddots  &  \ddots  &  \ddots  &  \vdots\\
\infty  &   \dots  &  \dots  &  \dots  &   \frac{1}{a^{r-1}}^{\textcolor{red}{@}}  &  \varepsilon^{\textcolor{red}{@}}   &  \infty  &  \dots  &  \dots  &   \dots  &  \vdots  &  \ddots  &  \ddots  &   \ddots  &  \ddots  &  \ddots  &   \ddots  &  \ddots  &  \vdots\\
\infty  &   \dots  &  \dots  &  \dots  &   \frac{1}{a^{r-1}}  &  \infty   &  \varepsilon^{\textcolor{red}{@}}  &  \dots  &  \dots  &   \dots  &  \vdots  &  \ddots  &  \ddots  &   \ddots  &  \ddots  &  \ddots  &   \ddots  &  \ddots  &  \vdots\\
\infty  &   \dots  &  \dots  &  \dots  &   \dots  &  \dots  &  \dots  &  \frac{1}{a^{r}}^{\textcolor{red}{@}}  &   \dots  &  \dots  &  \vdots  &  \ddots  &   \ddots  &  \ddots  &  \ddots  &   \ddots  &  \ddots  &  \ddots  &\vdots\\
\vdots  &   \vdots  &  \vdots  &  \vdots  &   \vdots  &  \vdots  &  \vdots  &  \vdots  &   \ddots  &  \vdots  &  \vdots  &  \ddots  &   \ddots  &  \ddots  &  \ddots  &   \ddots  &  \ddots  &  \ddots  &\vdots\\
\infty  &   \dots  &  \dots  &  \dots  &   \dots  &  \dots  &  \dots  &  \dots  &   \dots  &  \frac{1}{a^{2r-1}}^{\textcolor{red}{@}}  &  \infty  &  \dots  &   \dots  &  \dots  &  \dots  &   \dots  &  \dots  &  \dots  &0^{\textcolor{red}{@}}\\
};
\mvline[dashed, black]{m}{8};
\mvline[dashed, black]{m}{11};
\end{tikzpicture}
\caption{In this case we achieve a lower bound on $M$'s approximation ratio when $k=0$.}
\label{second_app}
\end{figure}

In order to finish the proof of the theorem we need to find values of $b_1,...,b_n$ for which the approximation ratios in \eqref{eq1} and \eqref{equ2} are at least $1+a$,  this is done in the following lemma (Lemma ~\ref{same_app_ratio}).

\begin{lemma} \label{same_app_ratio}
For every $a<2$ there exists a number $r\in \mathbb{N}$ such that the expressions ratios given in \eqref{eq1} and \eqref{equ2} are at least $1+a$.  
\end{lemma}

\begin{proof}[Proof of Lemma \ref{same_app_ratio}]
Consider the expression given in \eqref{eq1}. Next, we find the values of $b_1,...,b_n$ for which this ratio is at least $1+a$.

\begin{equation*}
\begin{split}
     &\frac{\frac{1}{a^{k-1}}+\frac{1}{a^k}+{\max\{\frac{2}{a^{k}}-(b_i-\frac{1}{a^{k}}+\varepsilon), \frac{1}{a^{k}}\}}+b_{k+1}+...+b_r+\frac{1}{a^{r+1}}+...+\frac{1}{a^{2r}}}{\frac{1}{a^{k-1}}} \geq \\
    & 1+\frac{4}{a}-{b_k}\cdot a^{k-1} +{b_{k+1}}\cdot a^{k-1} + ...+ {b_{r}}\cdot a^{k-1}+\frac{1}{a^{r-k+2}}+...+\frac{1}{a^{2r-k+1}}
\end{split}
\end{equation*}

\begin{equation*}
\begin{split}
    1+a &= 1+\frac{4}{a}-{b_k}\cdot a^{k-1} +{b_{k+1}}\cdot a^{k-1} + ...+ {b_{r}}\cdot a^{k-1}+\frac{1}{a^{r-k+2}}+...+\frac{1}{a^{2r-k+1}}\\
        &\hspace{0.25\linewidth}\Longleftrightarrow\\
        a &= \frac{4}{a}-{b_k}\cdot a^{k-1} +{b_{k+1}}\cdot a^{k-1} + ...+ {b_{r}}\cdot a^{k-1} +a^{k-1}\cdot(\frac{1}{a^r\cdot (1-a)}\cdot(\frac{1}{a^r}-1))\\
        &\hspace{0.25\linewidth}\Longleftrightarrow\\
        b_k\cdot a^{k-1} &= -a+\frac{4}{a}+{b_{k+1}}\cdot a^{k-1} + ...+ {b_{r}}\cdot a^{k-1} +a^{k-1}\cdot(\frac{1}{a^r\cdot (1-a)}\cdot(\frac{1}{a^r}-1))\\
        &\hspace{0.25\linewidth}\Longleftrightarrow\\
\end{split}
\end{equation*}

\begin{equation}\label{bk_formula}
        b_k =- \frac{1}{a^{k-2}}+\frac{4}{a^k}+{b_{k+1}} + ...+ {b_{r}} +(\frac{1}{a^r\cdot (1-a)}\cdot(\frac{1}{a^r}-1))    
\end{equation}

For every $1 \leq k\leq r$ let $s(k)=b_{k+1}+...+b_r$, then $s(r)=0 $ and let $z=\frac{1}{a^r\cdot (1-a)}\cdot(\frac{1}{a^r}-1)$. Now, we can rewrite equation \eqref{bk_formula} as:

\begin{equation*}
    s(k-1)-s(k)=-\frac{1}{a^{k-2}}+\frac{4}{a^k}+s(k) +z\\
    \Longleftrightarrow
    s(k)=\frac{(s(k-1)+\frac{1}{a^{k-2}}-\frac{4}{a^k}-z)}{2}
\end{equation*} 

Solving the recurrence relation
\begin{equation*}
\begin{split}
    s(k)=&\frac{(s(k-1)+\frac{1}{a^{k-2}}-\frac{4}{a^k}-z)}{2}\\
    s(r)=& 0
\end{split}
\end{equation*}

yields:
\begin{equation}\label{b_k_final}
\begin{split}
    s(k) &= (a+2)(a^{-r}\cdot2^{r-k}-a^{-k})+z(2^{r-k}-1)\\
    b_k &= s(k-1)-s(k)= a^{-r} \, 2^{r-k}\left(z a^r+a+2\right)-a^{-k}(a^2+a-2)\\
    b_k &= a^{-r}\, 2^{r-k}\left(\frac{(1-a^r)}{a^r\cdot(1-a)}+a+2\right)-a^{-k}(a^2+a-2)
\end{split}
\end{equation} 

We analyze the second expression which is given in \eqref{equ2}. By definition: $s(0)= b_1+...+b_r$, we will show that the expression given in \eqref{equ2} is approaching $\infty$ when $r$ approaches $\infty$ for every $\sqrt{2}<a<2$, this will complete the proof of the lemma.

\begin{equation*}
        1+b_1+\dots+b_r+\frac{1}{a^{r+1}}+\dots+\frac{1}{a^{2r}}>b_1+\dots+b_r=s(0)\\
        =(a+2)\cdot(a^{-r}\cdot2^{r}-1)+z\cdot(2^{r}-1)
\end{equation*}

For a fixed $\sqrt{2}<a<2$ define the sequence $x_r=(a+2)\cdot(a^{-r}\cdot2^{r}-1)+z\cdot(2^{r}-1)$.

Now, we show that this sequence tends to infinity:

\begin{equation} \label{app_to_infty}
\begin{split}
        s(0)=&(a+2)\cdot(a^{-r}\cdot2^{r}-1)+z\cdot(2^{r}-1)\\
        =&(a+2)\cdot(a^{-r}\cdot2^{r}-1)+\frac{(1-a^r)}{a^{2r}\cdot(1-a)}\cdot(2^{r}-1)\\
        =& (a+2)\cdot\left(\left(\frac{2}{a}\right)^r-1\right)+\frac{2^r-{a^r}\cdot2^r-1+a^r}{a^{2r}\cdot(1-a)}\\
        =& (a+2)\cdot\left(\left(\frac{2}{a}\right)^r-1\right)+{\left(\frac{2}{a^2}\right)^{r}}\cdot\frac{1}{(1-a)}-{\left(\frac{2}{a}\right)^r}\cdot\frac{1}{(1-a)}-\frac{1}{a^{2r}\cdot(1-a)}+\frac{1}{a^{r}\cdot(1-a)}\\
        =& \underbrace{(a+2)\cdot\left(\left(\frac{2}{a}\right)^r\right)}_{\xrightarrow[r \to \infty]{} \infty}-a-2+\underbrace{\left(\frac{1}{1-a}\right)\cdot\left(\underbrace{\left(\frac{2}{a^2}\right)^{r}}_{\xrightarrow[r \to \infty]{} 0}-\underbrace{\frac{1}{a^{2r}}+\frac{1}{a^{r}}}_{\xrightarrow[r \to \infty]{} 0}\right)}_{\xrightarrow[r \to \infty]{} 0} +\underbrace{\underbrace{{\left(\frac{2}{a}\right)^r}}_{\xrightarrow[r \to \infty]{} \infty}\cdot\underbrace{\frac{1}{(a-1)}}_{>0}}_{\xrightarrow[r \to \infty]{} \infty}\\
\end{split}
\end{equation}



\end{proof}

\subsubsection{Thrid Part: Concluding the Proof of Theorem \ref{main_thm}}\label{third_part}

Lemma ~\ref{same_app_ratio} almost concludes the proof of Theorem ~\ref{main_thm} we are left with proving Lemma ~\ref{bk_is_atleat} which can now that we have the exact values of $b_1,...,b_n$ be proved. In Table ~\ref{approximation} there are some specific values of $a$ and $n,r$ for which the lower bound hold. This section also contains an example (Example ~\ref{example_of_main_instance}) of the instance $\mathcal{A}_n$ for a specific value of $r$. 

\begin{table}[H]
\centering
\begin{tabular}{|l|l|l|l|l|l|l|l|l|l|}
\hline
$n (r)$                   &10 (3)            &13 (4)      & 16 (5)      & 31 (10) &91 (30)     &109 (36)\\ \hline
approximation ratio   & 2.873        & 2.911  &2.932    &2.966 & 2.988  & 2.990  \\ \hline
\end{tabular}
\caption{\label{approximation} The lower bound of the approximation ratio by $r$, where the number of players is $n$ and the number of jobs is $7r$  (all numbers have three decimal digits of precision).}
 \end{table}
 
Now, after we have found the specific values of $b_1,...,n_n$ we can prove Lemma ~\ref{bk_is_atleat}. 

\begin{lemma} \label{bk_is_atleat}
For every $1 \leq k \leq r$ it holds that $b_k\geq \frac{1}{a^k}$, for the relevant values of $a$.
\end{lemma}

\begin{proof}[Proof of Lemma \ref{bk_is_atleat}]
From \eqref{b_k_final}, we need to show that:
\begin{align*}
            a^{-r}\cdot2^{r-k}\cdot(\frac{(1-a^r)}{a^r\cdot(1-a)}+a+2)-a^{-k}\cdot(a^2+a-2)& \geq \frac{1}{a^k}\\
            \Longleftrightarrow \hspace{0.3\linewidth} &\\ 
            a^{-r+k}\cdot2^{r-k}\cdot(\frac{(1-a^r)}{a^r\cdot(1-a)}+a+2)-(a^2+a-2) & \geq 1\\
\end{align*}

Then we will prove the second inequality.
\begin{align*}
            &a^{-r+k}\cdot2^{r-k}\cdot(\frac{(1-a^r)}{a^r\cdot(1-a)}+a+2)-(a^2+a-2)\\
            &=(\frac{2}{a})^{r-k}\cdot(\frac{(1-a^r)}{a^r\cdot(1-a)}+a+2)-a^2-a+2\\
            &\underbrace{ \geq}_{a<2} (\frac{2}{a})^{0}\cdot(\frac{(1-a^r)}{a^r\cdot(1-a)}+a+2)-a^2-a+2\\
            &=(\frac{(1-a^r)}{a^r\cdot(1-a)}+a+2)-a^2-a+2\\
            &=\underbrace{\frac{(1-a^r)}{a^r\cdot(1-a)}-a^2+4}_{=e_r}\\
\end{align*}
Observe that ${\{e_r\}}_{r=1}^\infty$ is a monotonically increasing sequence ($a>1$). Therefore, if $e_r\geq 1$ for some value of $a$, then also $e_p \geq 1$ for every $p\geq r$ (for the same value of a).
More specifically, for $r=3$, it holds that $e_3\leq 1$ for $0<a\leq 1.973$, when $r=4$ it holds that $e_4\leq 1$ for $0<a\leq 1.987$, and when $r=16$ it holds that $e_{16}\leq 1$ for $0<a\leq2$.
\end{proof}

\begin{example}[an example of the instance $\mathcal{A}_n$, for $r=3, n=10$]
\label{example_of_main_instance}
When $r=3, n=10$, we have that $a\approx1.873$ and the approximation ratio is of $2.873$, below is the resulting instance $\mathcal{A}_n$ (all numbers have three decimal digits of precision)).\\

\begin{center}
\begin{tikzpicture}
\matrix [matrix of math nodes,left delimiter=(,right delimiter=)] (m)
{
1.141  &   1.067  &  1.067  &  0.509  &   0.570  &  0.570  &  0.222  &  0.304  &   0.304  &  0.081  &  0.043  &  0.023  &   \dots\\
1  &   \varepsilon  &  \infty  &  \dots  &   \dots  &  \dots  &  \dots  &  \dots  &   \dots  &  \dots  &  \dots  &  \dots  &   \dots \\
1  &   \infty  &  \varepsilon  &  \dots  &   \dots  &  \dots  &  \dots  &  \dots  &   \dots  &  \dots  &  \dots  &  \dots  &   \dots\\
\dots  &   \dots  &  \dots  &  0.533  &   \varepsilon  &  \infty  &  \dots  &  \dots  &   \dots  &  \dots  &  \dots  &  \dots  &   \dots&\\
\dots  &   \dots  &  \dots  &  0.533  &   \infty  &  \varepsilon  &  \dots  &  \dots  &   \dots  &  \dots  &  \dots  &  \dots  &   \dots\\
\dots  &   \dots  &  \dots  &  \dots  &   \dots  &  \dots  &  0.285  &  \varepsilon  &   \infty  &  \dots  &  \dots  &  \dots  &   D_3\\
\dots  &   \dots  &  \dots  &  \dots  &   \dots  &  \dots  &  0.285  &  \infty  &   \varepsilon  &  \dots  &  \dots  &  \dots  &   \dots\\
\dots  &   \dots  &  \dots  &  \dots  &   \dots  &  \dots  &  \dots  &  \dots  &   \dots  &  0.152  &  \dots  &  \dots  &   \dots\\
\dots  &   \dots  &  \dots  &  \dots  &   \dots  &  \dots  &  \dots  &  \dots  &   \dots  &  \dots  &  0.081  &  \dots  &   \dots\\
\dots  &   \dots  &  \dots  &  \dots  &   \dots  &  \dots  &  \dots  &  \dots  &   \dots  &  \dots  &  \dots  &  0.043  &   \dots\\
};

\mvline[dashed, black]{m}{10};
\mvline[dashed, black]{m}{13};

\end{tikzpicture}
\end{center}
\end{example}

\section{Lower Bounds For Small Instances}\label{small_instances}
 
In this section we prove new lower bounds for instances with a small number of players and jobs. Recall that in the previous proofs we had  one dummy item. The idea is to reduce the number of dummy items. We achieve this by removing the dummy items for some set of players and we make them share an item that is \emph{potentially} a dummy item for each player in the set. The challenge is to make an this item fulfill its potential when needed, without hurting the approximation ratio (at least not too much).


Throughout this section we denote by $\varepsilon^+,\varepsilon^-, \varepsilon^{--}, \varepsilon^{---}$ values that are as small as we wish, such that $\varepsilon^+>>\varepsilon^->>\varepsilon^{--}>>\varepsilon^{---}$. Thus, we will treat $\frac{\varepsilon^+}{\varepsilon^-}, \frac{\varepsilon^-}{\varepsilon^{--}}, \frac{\varepsilon^{--}}{\varepsilon^{---}}$ as $\infty$. 
 
\subsection{2 Players and 2 Jobs}\label{2*2}

\begin{theorem}\label{2*2_thm}
Every truthful mechanism for the unrelated machine scheduling problem with two machines and two jobs has an approximation ratio of at least 2. 
\end{theorem}

Let $M$ be a truthful mechanism for the $2 \times 2$ case of the unrelated machine scheduling problem with an approximation ratio better than 2. Consider the instance:
\[
D=\begin{pmatrix}
1 &\varepsilon^-\\
1 &\varepsilon^+\\
\end{pmatrix}
\]

\begin{lemma}\label{lemma6}
If $M$ has a finite approximation ratio then it must allocate the second job to the first player.
\end{lemma}

\begin{proof}[Proof of Lemma \ref{lemma6}]
Assume by contradiction that $M$ has a finite approximation ratio and it allocates the second job to the second player. Then, there are two possible cases. In the first case, $M$ allocates the first job to the first player, and in the second case, $M$ allocates the first job to the second player in addition to the second job.

In the first case, reduce the first player's cost for the first job to 0. By Lemma \ref{lemma1}, $M$ allocates the same jobs to the first player i.e., the first player will get job 1 and player 2 will get job 2 (see Figure~\ref{2_2_1_lemma}). In this case the mechanism's makespan is $\varepsilon^+$ whereas the optimal makespan is $\varepsilon^-$ (this can be achieved when the first player gets both jobs). Thus, $M$ has an approximation ratio of $\frac{\varepsilon^+}{\varepsilon^-}$ which is as large as we want.

\begin{figure}[h!]
\centering
$\begin{pmatrix}
0^{\textcolor{blue}{*}\textcolor{red}{@}} &{\varepsilon^-}^{\textcolor{red}{@}}\\
1 &{\varepsilon^+}^{\textcolor{blue}{*}}\\
\end{pmatrix}$
\caption{This is the instance that result from the transition described in Lemma \ref{lemma6}, Case 1.
}
\label{2_2_1_lemma}
\end{figure}

In the second case, $M$ allocates both jobs to the second player. Reduce the cost of the second player for the first job to 0. By Lemma \ref{lemma1}, we have that the second player keeps both jobs. Thus, $M$ has a makespan of $\varepsilon^+$ whereas the optimal makespan is $\varepsilon^-$ (see Figure~\ref{2_2_2lemma}). In this case, $M$ has an approximation ratio which can be made arbitrarily large. 

\begin{figure}[h!]
\centering
$\begin{pmatrix}
1 &{\varepsilon^-}^{\textcolor{red}{@}}\\
0^{\textcolor{blue}{*}\textcolor{red}{@}} &{\varepsilon^+}^{\textcolor{blue}{*}}\\
\end{pmatrix}$
\caption{This is the instance that result from the transition described in Lemma \ref{lemma6}, Case 2.
}
\label{2_2_2lemma}
\end{figure}

\end{proof}

By Lemma \ref{lemma6}, we have that $M$ allocates the second job to the first player. There are two possible cases: the first is that the first player gets the first job and the second is that the second player gets the first job. 

\subsubsection*{Case 1: $M$ Allocates Job 1 to Player 1}

In this case, in the instance $D$, $M$ allocates both jobs to the first player. Increase the second player's cost for the second job to $\infty$. This effectively makes the second job a dummy job for player 1. By Lemma \ref{lemma1}, the second player's allocation does not change. Thus, the first player keeps both jobs. Increase the cost of the first player for the second job to 1. By Lemma \ref{lemma3}, the first player keeps both jobs and $M$ has a makespan of 2 whereas the optimal makespan is 1 (this can be achieved when the first player gets the second job and the second player gets the first job) (see Figure~\ref{2_2_case1}).   

\begin{figure}[h!]
\centering
$\begin{pmatrix}
1^{\textcolor{blue}{*}} &{\varepsilon^-}^{\textcolor{blue}{*}}\\
1 &\varepsilon^+\\
\end{pmatrix}
\xrightarrow{}
\quad
\begin{pmatrix}
1^{\textcolor{blue}{*}} &{\varepsilon^-}^{\textcolor{blue}{*}}\\
1 &\infty\\
\end{pmatrix}
\xrightarrow{}
\quad
\begin{pmatrix}
1^{\textcolor{blue}{*}} &1^{\textcolor{blue}{*}\textcolor{red}{@}}\\
1^{\textcolor{red}{@}} &\infty\\
\end{pmatrix}$
\caption{The transitions described in the analysis of the first case of the proof of theorem 2.}
\label{2_2_case1}
\end{figure}

\subsubsection*{Case 2: $M$ Allocates Job 1 to the Second Player}

In this case we reduce the second player's cost for the second job to $\varepsilon^{--}$ and reduce its cost for the first job by $\varepsilon'>\varepsilon^+-\varepsilon^{--}$. Denote this instance by $D_1$ (see Figure~\ref{D_1}).

\begin{figure}[h!]
\centering
$\begin{pmatrix}
1 &\varepsilon^-\\
1-\varepsilon' &\varepsilon^{--}\\
\end{pmatrix}$
\caption{The instance $D_1$}
\label{D_1}
\end{figure}

\begin{lemma}\label{lemma7}
In the instance $D_1$, $M$ must allocate the first job to the second player. 
\end{lemma}

\begin{proof}[Proof of Lemma \ref{lemma7}]
Consider the weak monotonicity inequality applied on the second player where $x'$ is the allocation of $M$ in the instance $D_1$ and assume by contradiction that ${x'_2}^1=0$:
\begin{align*}
    &(1-(1-\varepsilon'))\cdot(1-{x'_2}^1)+(\varepsilon^+-\varepsilon^{--})\cdot(0-{x_2}^2)\\
    & = (\varepsilon')+(\varepsilon^+-\varepsilon^{--})\cdot(0-{x_2}^2)\\
    & \leq 0
\end{align*}
In this case the inequality does not hold since $\varepsilon'>\varepsilon^+-\varepsilon^{--}$.
\end{proof}

By Lemma \ref{lemma7}, we have that in $D_1$ the second player gets the first job. Similarly to Lemma \ref{lemma6}, we can show that $M$ allocates the second job to the second player in $D_1$ (otherwise we can decrease the second player's cost for the first job to 0. Using Lemma \ref{lemma1}, we get that $M$ has an approximation ratio of $\frac{\varepsilon^-}{\varepsilon^{--}}$ which is as large as we want).

It remains to show that in $D_1$, if the second player gets both jobs then $M$ has an approximation ratio of 2. 
In this case, increase the first player's cost for the second job to $\infty$. By Lemma \ref{lemma1}, the first player will not get either job. Thus, the second player will keep both jobs and the makespan of $M$ will be $2-\varepsilon'$, whereas the optimal makespan is 1 (this can be achieved when the first player gets the first job and the second player gets the second job). See Figure~\ref{2_2_end}.

\begin{figure}[h!]
\centering
$\begin{pmatrix}
1 &\varepsilon^-\\
(1-\varepsilon')^{\textcolor{blue}{*}} &{\varepsilon^{--}}^{\textcolor{blue}{*}}\\
\end{pmatrix}
\xrightarrow{}
\quad
\begin{pmatrix}
1&\infty\\
(1-\varepsilon')^{\textcolor{blue}{*}} &{\varepsilon^{--}}^{\textcolor{blue}{*}}\\
\end{pmatrix}
\xrightarrow{}
\quad
\begin{pmatrix}
{1}^{\textcolor{red}{@}} &\infty\\
(1-\varepsilon')^{\textcolor{blue}{*}} &1^{\textcolor{blue}{*}\textcolor{red}{@}}\\
\end{pmatrix}$
\caption{The transitions described in the case that in $D_1$ the second player gets both jobs (as analyzed in the second case of the proof of theorem 2).
}
\label{2_2_end}
\end{figure}

\subsection{3 Players and 3 Jobs}\label{3*3}

\begin{theorem}\label{3*3_thm}
Every truthful mechanism for the unrelated machine scheduling with three machines and three jobs has an approximation ratio of at least $2.2055$ 
\end{theorem}

Let $M$ be a truthful mechanism for the $3\times3$ case of the unrelated machine scheduling problem with an approximation ratio better than 2.2055. Consider the instance:
\[
E=\begin{pmatrix}
\infty &c      &\varepsilon^+\\
b      &\infty & \infty\\
a      &b      &\varepsilon^-
\end{pmatrix}
\]

We will show that $M$ has an approximation ratio of at least $\min\{1+\frac{c}{b},\frac{b}{a},\frac{a+b+c}{c}\}$ where $a<b<c$. We can then prove Theorem \ref{3*3_thm} by choosing $a=1$, $b\approx2.2055$ and $c\approx 2.6589$.

We divide the analysis to four cases based on the allocation of $M$ in $E$. In the first case, $M$ allocates the second job to the first player. In the second, $M$ allocates the first job to the second player and the second job to the third player.  In the third case, $M$ allocates the first three jobs to the third player. In the forth case, $M$ allocates the first two jobs to the third player and the third job to the first player.

\subsubsection*{Case 1: $M$ Allocates the Second Job to the First Player}
There are two possible cases for the allocation of the third job in $E$. In the first case, $M$ allocates the third job to the first player and in the second case, $M$ allocates it to the third player (otherwise the approximation ratio can be made arbitrarily large).  

\subsubsection*{Case 1.1: $M$ Allocates the Third Job to the First Player}
In this case, in the instance $E$, $M$ allocates the second and third jobs to the first player. Increase the third player's cost for the third job to $\infty$ which makes the third job a dummy job for the first player. By Lemma \ref{lemma1} we have that the third player does not get the second and third jobs. Thus, since $M$ has a finite approximation ratio, the first player keeps these jobs. Increase the cost of the the first player for his dummy job to $b$. By Lemma \ref{lemma3}, the first player gets the second and third jobs which results in an approximation ratio of $\frac{c+b}{b}$, as is visualized below.

\[
\begin{pmatrix}
\infty &c^{\textcolor{blue}{*}}      &{\varepsilon^+}^{\textcolor{blue}{*}}\\
b      &\infty & \infty\\
a      &b      &{\varepsilon^-}
\end{pmatrix}
\xrightarrow{}
\quad
\begin{pmatrix}
\infty &c^{\textcolor{blue}{*}}      &{\varepsilon^+}^{\textcolor{blue}{*}}\\
b      &\infty & \infty\\
a      &b      &\infty
\end{pmatrix}
\xrightarrow{}
\quad
\begin{pmatrix}
\infty &c^{\textcolor{blue}{*}}      &{b}^{\textcolor{blue}{*}\textcolor{red}{@}}\\
b^{\textcolor{red}{@}}      &\infty & \infty\\
a      &b^{\textcolor{red}{@}}      &\infty
\end{pmatrix}
\\
\]

\subsubsection*{Case 1.2: $M$ does not Allocate the Third Job to the First Player}
In this case, in $E$, $M$ allocates the second job but not the third job to the first player. Reduce the first player's cost for the third job to $\varepsilon^{--}$ and reduce his cost for the second job by $\varepsilon'>\varepsilon^+-\varepsilon^{--}$. Denote this instance by $E_1$ (as is visualized below).

\[
\begin{pmatrix}
\infty &c^{\textcolor{blue}{*}}      &\varepsilon^+\\
b      &\infty & \infty\\
a      &b      &{\varepsilon^-}^{\textcolor{blue}{*}}
\end{pmatrix}
\xrightarrow{}
\quad
\underbrace{\begin{pmatrix}
\infty &c-\varepsilon'      &\varepsilon^{--}\\
b      &\infty & \infty\\
a      &b      &\varepsilon^-
\end{pmatrix}}_{=E_1}
\\
\]

\begin{lemma}\label{lemma8}
Suppose that in $E$, $M$ allocates the second job, but not the third job, to the first player. Then, in $E_1$, $M$ allocates the second job to the first player. 
\end{lemma}

\begin{proof}[Proof of Lemma \ref{lemma8}]
Consider the weak monotonicity inequality applied on the first player where $x'$ is the allocation of $M$ in $E_1$ and $x$ is the allocation of $M$ in $E$. Assume by contradiction that ${x'_1}^2=0$:
\begin{align*}
    &(\infty-\infty)\cdot(0-{x_1}^1)+(c-(c-\varepsilon'))\cdot(1-{x'_1}^2)+(\varepsilon^+-\varepsilon^{--})\cdot(0-{x'_1}^3)\\
    & = (\varepsilon')+(\varepsilon^+-\varepsilon^{--})\cdot(0-{x'_1}^3)\\
    & \leq 0
\end{align*}
and we arrive at a contradiction, since $\varepsilon'>\varepsilon^+-\varepsilon^{--}$.
\end{proof}

By Lemma \ref{lemma8}, in $E_1$, $M$ allocates the second job to the first player. There are two possible cases in the instance $E_1$: the first is that $M$ allocates the third job to the third player. The second case is that $M$ allocates the third job to the first player.

\subsubsection*{Case 1.2.1} In this case, in $E_1$, $M$ allocates the third job to the third player and the second job to the first player. Reduce the cost of the first player for the second job to 0, and by Lemma \ref{lemma1}, the first player still gets only the second job (as visualized below).

\[
\begin{pmatrix}
\infty &{c-\varepsilon'}^{\textcolor{blue}{*}}      &\varepsilon^{--}\\
b      &\infty & \infty\\
a      &b      &{\varepsilon^-}^{\textcolor{blue}{*}} 
\end{pmatrix}
\underbrace{\xrightarrow{}}_{\text{By Lemma \ref{lemma1}}}
\quad
\begin{pmatrix}
\infty &{0}^{\textcolor{blue}{*}}      &\varepsilon^{--}\\
b      &\infty & \infty\\
a      &b      &{\varepsilon^-}^{\textcolor{blue}{*}} 
\end{pmatrix}
\]

There are again two possible cases. In the first, the second player is assigned the first job, which results in an approximation ratio of $\frac{b}{a}$ (the makespan of $M$ is $b$ whereas the optimal one is $a$ which can be achieved, for example, when the first player takes the second and third jobs and the third player takes the first job). 

\[
\begin{pmatrix}
\infty &0^{\textcolor{blue}{*}\textcolor{red}{@}}      &{\varepsilon^{--}}^{\textcolor{red}{@}}\\
b^{\textcolor{blue}{*}}      &\infty & \infty\\
a^{\textcolor{red}{@}}      &b      &{\varepsilon^-}^{\textcolor{blue}{*}} 
\end{pmatrix}
\]
In the second case, the third player takes the first and third jobs and the first player takes the second job. Reduce the third player's cost for the first job to 0, and by Lemma \ref{lemma1}, we have that the third player will still get the first and third jobs. Then, $M$ has a makespan of $\varepsilon^-$ whereas the optimal one is $\varepsilon^{--}$ (this can be achieved when the first player gets the second and third jobs and the third player gets the first job) which results in an approximation ratio of $\frac{\varepsilon^-}{\varepsilon^{--}}$, which can be made arbitrarily large.  

\[
\begin{pmatrix}
\infty &{0}^{\textcolor{blue}{*}}      &{\varepsilon^{--}}\\
b      &\infty & \infty\\
a^{\textcolor{blue}{*}}      &b      &{\varepsilon^-}^{\textcolor{blue}{*}} 
\end{pmatrix}
\underbrace{\xrightarrow{}}_{\text{By Lemma \ref{lemma1}}}
\quad
\begin{pmatrix}
\infty &{0}^{\textcolor{blue}{*}\textcolor{red}{@}}      &{\varepsilon^{--}}^{\textcolor{red}{@}}\\
b      &\infty & \infty\\
0^{\textcolor{blue}{*}\textcolor{red}{@}}     &b      &{\varepsilon^-}^{\textcolor{blue}{*}} 
\end{pmatrix}
\]

\subsubsection*{Case 1.2.2} In this case, in $E_1$, $M$ allocates the second and third jobs to the first player.
\[
\begin{pmatrix}
\infty &{c-\varepsilon'}^{\textcolor{blue}{*}}      &{\varepsilon^{--}}^{\textcolor{blue}{*}}\\
b      &\infty & \infty\\
a      &b      &\varepsilon^-
\end{pmatrix}
\]

The analysis is very similar to Case 1.1. Increase the third player's cost for the third job to $\infty$, which makes the third job a dummy job for the first player. By Lemma \ref{lemma1}, we have that the third player does not get the second or the third jobs. Thus, since $M$ has a finite approximation ratio the first player keeps these jobs. Increase the first player's dummy job's cost to $b$. By Lemma \ref{lemma3}, the first player gets the second and third jobs which results in an approximation ratio of $\frac{c+b}{b}$. This is visualized below.

\[
\begin{pmatrix}
\infty &{c-\varepsilon'}^{\textcolor{blue}{*}}      &{\varepsilon^{--}}^{\textcolor{blue}{*}}\\
b      &\infty & \infty\\
a      &b      &\varepsilon^-
\end{pmatrix}
\underbrace{\xrightarrow{}}_{\text{By Lemma \ref{lemma1}}}
\quad
\begin{pmatrix}
\infty &{c-\varepsilon'}^{\textcolor{blue}{*}}      &{\varepsilon^{--}}^{\textcolor{blue}{*}}\\
b      &\infty & \infty\\
a      &b      &\infty
\end{pmatrix}
\underbrace{\xrightarrow{}}_{\text{By Lemma \ref{lemma3}}}
\quad
\begin{pmatrix}
\infty &{c-\varepsilon'}^{\textcolor{blue}{*}}      &{b}^{\textcolor{blue}{*}\textcolor{red}{@}}\\
b^{\textcolor{red}{@}}      &\infty & \infty\\
a      &b^{\textcolor{red}{@}}      &\infty
\end{pmatrix}
\]

\subsubsection*{Case 2: $M$ Allocates the First Job to the Second Player and the Second Job to the Third Player}

In this case, $M$ allocates the first job to the second player and the second job to the third player in $E$. Reduce the cost of the third player for the second job to 0. By Lemma \ref{lemma1}, the third player will not get the first job and since $M$ has a finite approximation ratio, the second player will get it. Then, $M$ will have an approximation ratio of $\frac{b}{a}$. This is visualized below. 

\[
\begin{pmatrix}
\infty &c      &{\varepsilon^+}\\
b^{\textcolor{blue}{*}}      &\infty & \infty\\
a      &b^{\textcolor{blue}{*}}      &{\varepsilon^-}
\end{pmatrix}
\underbrace{\xrightarrow{}}_{\text{By Lemma \ref{lemma1}}}
\quad
\begin{pmatrix}
\infty &c      &{\varepsilon^+}\\
b^{\textcolor{blue}{*}}      &\infty & \infty\\
a^{\textcolor{red}{@}}      &0^{\textcolor{blue}{*}\textcolor{red}{@}}      &{\varepsilon^-}^{\textcolor{red}{@}}
\end{pmatrix}
\]

\subsubsection*{Case 3: $M$ Allocates All Three Jobs to the Third Player}

In this case, $M$ allocates all three jobs to the third player. Increase the cost of the first player for the third job to $\infty$. This makes the third job a dummy job for the first player. By Lemma \ref{lemma1}, the first player will not get any job. Denote the resulting instance by $E_2$.
\[
\begin{pmatrix}
\infty &c      &{\varepsilon^+}\\
b      &\infty & \infty\\
a^{\textcolor{blue}{*}}      &b^{\textcolor{blue}{*}}      &{\varepsilon^-}^{\textcolor{blue}{*}}
\end{pmatrix}
\underbrace{\xrightarrow{}}_{\text{By Lemma \ref{lemma1}}}
\quad
\underbrace{\begin{pmatrix}
\infty &c      &\infty\\
b      &\infty & \infty\\
a      &b^{\textcolor{blue}{*}}      &{\varepsilon^-}^{\textcolor{blue}{*}}
\end{pmatrix}}_{=E_2}
\]

There are two possible cases based on the allocation of $M$ in the instance $E_2$. In the first case, $M$ allocates the first job to the second player and in the second case, $M$ allocates the first job to the third player. 

\subsubsection*{Case 3.1} In this case, $M$ allocates the first job to the second player and allocates the second and third jobs to the third player in the instance $E_2$. Reduce the third player cost for the second job to 0 and by Lemma \ref{lemma1}, the third player will not get the first job. Thus, since $M$ has a finite approximation ratio, the second player will get the first job which results in an approximation ratio of $\frac{b}{a}$. This is shown below.

\[
\begin{pmatrix}
\infty &c      &\infty\\
b^{\textcolor{blue}{*}}      &\infty & \infty\\
a      &b^{\textcolor{blue}{*}}      &{\varepsilon^-}^{\textcolor{blue}{*}}
\end{pmatrix}
\underbrace{\xrightarrow{}}_{\text{By Lemma \ref{lemma1}}}
\quad
\underbrace{\begin{pmatrix}
\infty &c      &\infty\\
b^{\textcolor{blue}{*}}      &\infty & \infty\\
a^{\textcolor{red}{@}}      &0^{\textcolor{blue}{*}\textcolor{red}{@}}      &{\varepsilon^-}^{\textcolor{blue}{*}\textcolor{red}{@}}
\end{pmatrix}}_{=E_2}
\]

\subsubsection*{Case 3.2} In this case, $M$ allocates all three jobs to the third player in the instance $E_2$. Increase the cost of the third player for its dummy job (the third job) to $c$ and by Lemma \ref{lemma3}, the third player will get all three jobs. In this case $M$ has an approximation ratio of $\frac{a+b+c}{c}$, as depicted below. 

\[
\begin{pmatrix}
\infty &c      &\infty\\
b      &\infty & \infty\\
a^{\textcolor{blue}{*}}      &b^{\textcolor{blue}{*}}      &{\varepsilon^-}^{\textcolor{blue}{*}}
\end{pmatrix}
\underbrace{\xrightarrow{}}_{\text{By Lemma \ref{lemma3}}}
\quad
\begin{pmatrix}
\infty &c^{\textcolor{red}{@}}      &\infty\\
b^{\textcolor{red}{@}}      &\infty & \infty\\
a^{\textcolor{blue}{*}}      &b^{\textcolor{blue}{*}}      &{c}^{\textcolor{blue}{*}\textcolor{red}{@}}
\end{pmatrix}
\]

\subsubsection*{Case 4: $M$ Allocates the First Two Jobs to the Third Player and the Third job to the First Player}
Reduce the third player's cost for the first two jobs to 0. By Lemma \ref{lemma1}, the third player will not get the third job and since the approximation is finite the third player will.
In this case, $M$ has a makespan of $\varepsilon^+$ whereas the optimal makespan is of $\varepsilon^-$ which results in an approximation ratio of $\frac{\varepsilon^+}{\varepsilon^-}$ which can be arbitrarily large. 

\[
\begin{pmatrix}
\infty &c      &{\varepsilon^+}^{\textcolor{blue}{*}}\\
b      &\infty & \infty\\
a^{\textcolor{blue}{*}}      &b^{\textcolor{blue}{*}}      &{\varepsilon^-}
\end{pmatrix}
\underbrace{\xrightarrow{}}_{\text{By Lemma \ref{lemma1}}}
\quad
\underbrace{\begin{pmatrix}
\infty &c      &{\varepsilon^+}^{\textcolor{blue}{*}}\\
b      &\infty & \infty\\
0^{\textcolor{blue}{*}\textcolor{red}{@}}      &0^{\textcolor{blue}{*}\textcolor{red}{@}}      &{\varepsilon^-}^{\textcolor{red}{@}}
\end{pmatrix}}_{=E_2}
\]

\section{3 Players and 4 Jobs}\label{3*4}

\begin{theorem}\label{3_4_thm}
Every truthful mechanism for the unrelated machine scheduling with three machines and four jobs has an approximation ratio of at least $1+\sqrt{2}$. 
\end{theorem}

Let $M$ be a truthful mechanism for the $3\times 4$ case of the unrelated machine scheduling problem with an approximation ratio better than $1+\sqrt{2}$.

 Consider the instance:
\[
F=\begin{pmatrix}
\infty &x               &\infty               &\varepsilon^{---}\\
1      &\varepsilon^{-} &\varepsilon^{-}      &\infty\\
1      &\varepsilon^{+} &\varepsilon^{--}     &\infty\\
\end{pmatrix}
\]

We will prove that $M$ has an approximation ratio of at least $\min \{1+x,\frac{2+x}{x}\}$ where $x>1$. This proves Theorem \ref{3_4_thm} when using $x=\sqrt{2}$.  

Consider the instance $F$. We divide the analysis to cases based on the allocation of $M$. In the first case, $M$ allocates the second job to the first player. In the second case, $M$ allocates the first two jobs to the second player. In the third case, $M$ allocates the first job to the second player and the second job to the third player. In the fourth case, $M$ allocates the first job to the third player and the second job to the second player. In the fifth case, $M$ allocates the first two jobs to the third player.

\subsubsection*{Case 1} In this case, in the instance $F$, $M$ allocates the second job to the first player. Increase the first player's dummy job's cost to 1 and by Lemma \ref{lemma3} we have that the first player keeps the second and forth jobs which results in a makespan of $1+x$ whereas the optimal makespan is 1. 

\[
\begin{pmatrix}
\infty &x^{\textcolor{blue}{*}}               &\infty               &{\varepsilon^{---}}^{\textcolor{blue}{*}}\\
1      &\varepsilon^{-} &\varepsilon^{-}      &\infty\\
1      &\varepsilon^{+} &\varepsilon^{--}     &\infty\\
\end{pmatrix}
\underbrace{\xrightarrow{}}_{\text{By Lemma \ref{lemma3}}}
\quad
\begin{pmatrix}
\infty &x^{\textcolor{blue}{*}}               &\infty               &1^{\textcolor{blue}{*}\textcolor{red}{@}}\\
1^{\textcolor{red}{@}}      &{\varepsilon^{-}}^{\textcolor{red}{@}} &\varepsilon^{-}      &\infty\\
1      &\varepsilon^{+} &{\varepsilon^{--}}^{\textcolor{red}{@}}     &\infty\\
\end{pmatrix}
\]

\subsubsection*{Case 2}

In this case, in the instance $F$, $M$ allocates the first two jobs to the second player. 
\[
\begin{pmatrix}
\infty &x               &\infty             &{\varepsilon^{---}}\\
1^{\textcolor{blue}{*}}      &{\varepsilon^{-}}^{\textcolor{blue}{*}} &\varepsilon^{-}      &\infty\\
1      &\varepsilon^{+} &\varepsilon^{--}     &\infty\\
\end{pmatrix}
\underbrace{\xrightarrow{}}_{\text{By Lemma \ref{lemma1}}}
\quad
\begin{pmatrix}
\infty &x               &\infty             &{\varepsilon^{---}}\\
1^{\textcolor{blue}{*}}      &{\varepsilon^{-}} &\varepsilon^{-}      &\infty\\
1      &\infty &\varepsilon^{--}     &\infty\\
\end{pmatrix}
\]
By Lemma \ref{lemma1}, the third player doesn't get the first or the second jobs. In the case that the first player gets the second job we have an $1+x$ approximation ratio as shown below:
\[
\begin{pmatrix}
\infty &x^{\textcolor{blue}{*}}               &\infty             &{\varepsilon^{---}}^{\textcolor{blue}{*}}\\
1^{\textcolor{blue}{*}}      &{\varepsilon^{-}} &\varepsilon^{-}      &\infty\\
1      &\infty &\varepsilon^{--}     &\infty\\
\end{pmatrix}
\underbrace{\xrightarrow{}}_{\text{By Lemma \ref{lemma3}}}
\quad
\begin{pmatrix}
\infty &x^{\textcolor{blue}{*}}               &\infty             &{1}^{\textcolor{blue}{*}\textcolor{red}{@}}\\
1     &{\varepsilon^{-}}^{\textcolor{red}{@}}  &{\varepsilon^{-}}^{\textcolor{red}{@}}       &\infty\\
1^{\textcolor{red}{@}}       &\infty &\varepsilon^{--}     &\infty\\
\end{pmatrix}
\]

Otherwise, the second player gets the first two jobs.
\[
\begin{pmatrix}
\infty &x               &\infty             &{\varepsilon^{---}}\\
1^{\textcolor{blue}{*}}      &{\varepsilon^{-}}^{\textcolor{blue}{*}} &\varepsilon^{-}      &\infty\\
1      &\infty &\varepsilon^{--}     &\infty\\
\end{pmatrix}
\xrightarrow{}
\quad
\underbrace{\begin{pmatrix}
\infty &x               &\infty             &{\varepsilon^{---}}\\
1      &1 &{\varepsilon^{-}}     &\infty\\
1      &\infty &\varepsilon^{--}     &\infty\\
\end{pmatrix}}_{=F_1}
\]
There are two possible cases. In the first, the first player gets the second job in the instance $F_1$ (this is analyzed in Case 2.1). In he second case, the second player keeps the second job (this analyzed in Case 2.2).
\subsubsection*{Case 2.1}
\[
\begin{pmatrix}
\infty &x^{\textcolor{blue}{*}}               &\infty             &{\varepsilon^{---}}\\
1      &1 &\varepsilon^{-}      &\infty\\
1      &\infty &\varepsilon^{--}     &\infty\\
\end{pmatrix}
\underbrace{\xrightarrow{}}_{\text{By Lemma \ref{lemma3}}}
\quad
\begin{pmatrix}
\infty &x^{\textcolor{blue}{*}}               &\infty             &1^{\textcolor{blue}{*}\textcolor{red}{@}}\\
1      &1^{\textcolor{red}{@}} &{\varepsilon^{-}}^{\textcolor{red}{@}}      &\infty\\
1^{\textcolor{red}{@}}      &\infty &\varepsilon^{--}     &\infty\\
\end{pmatrix}
\]
In this case, $M$ has an approximation ratio of $1+x$.

\subsubsection*{Case 2.2}In this case, in the instance $F$, $M$ allocates the first two jobs to the second player and in the instance $F_1$, $M$ allocates the second job to the second player. By Lemma \ref{lemma4}, for $i=2, j_1=1, j_2=2$ we have that in the instance $F_1$ the second player gets the first two jobs. 

There are two possible cases, in the first case the second player gets the first three jobs in $F_1$ (analyzed in Case 2.2.1). In the second case the second player gets the first two jobs and the third player gets the third job (analyzed in Case 2.2.2).

\subsubsection*{Case 2.2.1}
\[
\begin{pmatrix}
\infty &x               &\infty             &{\varepsilon^{---}}\\
1^{\textcolor{blue}{*}}      &1^{\textcolor{blue}{*}} &{\varepsilon^{-}}^{\textcolor{blue}{*}}      &\infty\\
1      &\infty &\varepsilon^{--}     &\infty\\
\end{pmatrix}
\underbrace{\xrightarrow{}}_{\text{By Lemma \ref{lemma1}}}
\quad
\begin{pmatrix}
\infty &x               &\infty             &{\varepsilon^{---}}^{\textcolor{red}{@}} \\
0^{\textcolor{blue}{*}\textcolor{red}{@}}      &0^{\textcolor{blue}{*}\textcolor{red}{@}} &{\varepsilon^{-}}^{\textcolor{blue}{*}}      &\infty\\
1      &\infty &{\varepsilon^{--}}^{\textcolor{red}{@}}     &\infty\\
\end{pmatrix}
\]

This results in an approximation ratio of $\frac{\varepsilon^{-}}{\varepsilon^{--}}$ which can be made arbitrarily large.

\subsubsection*{Case 2.2.2}
Let $\varepsilon'>\varepsilon^{-}-\varepsilon^{---}$. 
\[
\begin{pmatrix}
\infty &x               &\infty             &{\varepsilon^{---}}\\
1^{\textcolor{blue}{*}}      &1^{\textcolor{blue}{*}} &{\varepsilon^{-}}      &\infty\\
1      &\infty &{\varepsilon^{--}}^{\textcolor{blue}{*}}     &\infty\\
\end{pmatrix}
\xrightarrow{}
\quad
\underbrace{\begin{pmatrix}
\infty &x               &\infty             &{\varepsilon^{---}}\\
{1-\varepsilon'}      &{1-\varepsilon'} &{\varepsilon^{---}}      &\infty\\
1      &\infty &\varepsilon^{--}    &\infty\\
\end{pmatrix}}_{=F_2}
\]
Similarly to Lemma \ref{lemma8}, it can be shown that the second player keeps the first two jobs in the instance $F_2$. There are two possible cases. In the first, the second player gets the first three jobs in the instance $F_2$ (analyzed in Case 2.2.2.1). In the second case the second player gets the first two jobs and the third player gets the third job (analyzed in Case 2.2.2.2). 

\subsubsection*{Case 2.2.2.1}
\[
\begin{pmatrix}
\infty &x               &\infty             &{\varepsilon^{---}}\\
{1-\varepsilon'}^{\textcolor{blue}{*}}      &{1-\varepsilon'}^{\textcolor{blue}{*}} &{\varepsilon^{---}}^{\textcolor{blue}{*}}     &\infty\\
1      &\infty &\varepsilon^{--}     &\infty\\
\end{pmatrix}
\underbrace{\xrightarrow{}}_{\text{By Lemma \ref{lemma1}}}
\quad
\begin{pmatrix}
\infty &x               &\infty             &{\varepsilon^{---}}\\
{1-\varepsilon'}^{\textcolor{blue}{*}}      &{1-\varepsilon'} &{\varepsilon^{---}}^{\textcolor{blue}{*}}     &\infty\\
1      &\infty &\infty     &\infty\\
\end{pmatrix}
\]

By Lemma \ref{lemma1}, the third player doesn't get a job from the first three jobs. There are two possible cases, the first is that the first player takes the second job (Case 2.2.2.1.1) and the second is that the second player takes the second job (and jobs 1,3) (Case 2.2.2.1.2). 

\subsubsection*{Case 2.2.2.1.1}
\[
\begin{pmatrix}
\infty &x^{\textcolor{blue}{*}}               &\infty             &{\varepsilon^{---}}^{\textcolor{blue}{*}}\\
{1-\varepsilon'}^{\textcolor{blue}{*}}      &{1-\varepsilon'} &{\varepsilon^{---}}^{\textcolor{blue}{*}}     &\infty\\
1      &\infty &\infty     &\infty\\
\end{pmatrix}
\underbrace{\xrightarrow{}}_{\text{By Lemma \ref{lemma3}}}
\quad
\begin{pmatrix}
\infty &x^{\textcolor{blue}{*}}               &\infty             &{1}^{\textcolor{blue}{*}\textcolor{red}{@}}\\
{1-\varepsilon'}    &{1-\varepsilon'}^{\textcolor{red}{@}} &{\varepsilon^{---}}^{\textcolor{red}{@}}     &\infty\\
1^{\textcolor{red}{@}}      &\infty &\infty     &\infty\\
\end{pmatrix}
\]

In this case, $M$ has an approximation ratio of $1+x$

\subsubsection*{Case 2.2.2.1.2}
\[
\begin{pmatrix}
\infty &x               &\infty             &{\varepsilon^{---}}^{\textcolor{blue}{*}}\\
{1-\varepsilon'}^{\textcolor{blue}{*}}      &{1-\varepsilon'}^{\textcolor{blue}{*}} &{\varepsilon^{---}}^{\textcolor{blue}{*}}     &\infty\\
1      &\infty &\infty     &\infty\\
\end{pmatrix}
\underbrace{\xrightarrow{}}_{\text{By Lemma \ref{lemma3}}}
\quad
\begin{pmatrix}
\infty &x^{\textcolor{red}{@}}               &\infty             &{\varepsilon^{---}}^{\textcolor{blue}{*}\textcolor{red}{@}}\\
{1-\varepsilon'}^{\textcolor{blue}{*}}      &{1-\varepsilon'}^{\textcolor{blue}{*}} &{x}^{\textcolor{blue}{*}\textcolor{red}{@}}     &\infty\\
1^{\textcolor{red}{@}}      &\infty &\infty     &\infty\\
\end{pmatrix}
\]

In this case, $M$ has an approximation ratio of $\frac{2+x}{x}$. 

\subsubsection*{Case 2.2.2.2}
\[
\begin{pmatrix}
\infty &x               &\infty             &{\varepsilon^{---}}\\
{1-\varepsilon'}^{\textcolor{blue}{*}}      &{1-\varepsilon'}^{\textcolor{blue}{*}} &{\varepsilon^{---}}     &\infty\\
1      &\infty &{\varepsilon^{--}}^{\textcolor{blue}{*}}     &\infty\\
\end{pmatrix}
\underbrace{\xrightarrow{}}_{\text{By Lemma \ref{lemma1}}}
\quad
\begin{pmatrix}
\infty &x               &\infty             &{\varepsilon^{---}}^{\textcolor{red}{@}}\\
{0}^{\textcolor{blue}{*}\textcolor{red}{@}}      &{0}^{\textcolor{blue}{*}\textcolor{red}{@}} &{\varepsilon^{---}}^{\textcolor{red}{@}}      &\infty\\
1      &\infty &{\varepsilon^{--}}^{\textcolor{blue}{*}}    &\infty\\
\end{pmatrix}
\]

By Lemma \ref{lemma1}, the second player doesn't get the third job and since $M$ has a finite approximation ratio the third job goes to the third player. This results in an approximation ratio of $\frac{\varepsilon^{--}}{\varepsilon^{---}}$ which can be made arbitrarily large.

\subsubsection*{Case 3}
In this case, $M$ has an approximation ratio of $\frac{\varepsilon+}{\varepsilon^-}$ which can be made arbitrarily large. 
\[
\begin{pmatrix}
\infty &x              &\infty               &{\varepsilon^{---}}\\
1^{\textcolor{blue}{*}}       &\varepsilon^{-} &\varepsilon^{-}      &\infty\\
1      &{\varepsilon^{+}}^{\textcolor{blue}{*}}  &\varepsilon^{--}     &\infty\\
\end{pmatrix}
\underbrace{\xrightarrow{}}_{\text{By Lemma \ref{lemma1}}}
\quad
\begin{pmatrix}
\infty &x              &\infty               &{\varepsilon^{---}}^{\textcolor{red}{@}} \\
0^{{\textcolor{blue}{*}}\textcolor{red}{@}}       &{\varepsilon^{-}}^{\textcolor{red}{@}} &{\varepsilon^{-}}     &\infty\\
1      &{\varepsilon^{+}}^{\textcolor{blue}{*}}  &{\varepsilon^{--}}^{\textcolor{red}{@}}      &\infty\\
\end{pmatrix}
\]

\subsubsection*{Case 4}
Let $\varepsilon'>\varepsilon^+-\varepsilon^{--}$
\[
\begin{pmatrix}
\infty &x              &\infty               &{\varepsilon^{---}}^{\textcolor{blue}{*}}\\
1      &{\varepsilon^{-}}^{\textcolor{blue}{*}}  &\varepsilon^{-}      &\infty\\
1^{\textcolor{blue}{*}}       &\varepsilon^{+} &\varepsilon^{--}     &\infty\\
\end{pmatrix}
\xrightarrow{}
\quad
\underbrace{\begin{pmatrix}
\infty &x              &\infty               &{\varepsilon^{---}}\\
1      &{\varepsilon^{-}}  &\varepsilon^{-}      &\infty\\
{1-\varepsilon'}^{\textcolor{blue}{*}}        &\varepsilon^{--} &\varepsilon^{--}     &\infty\\
\end{pmatrix}}_{=F_3}
\]

Similarly to Lemma \ref{lemma8}, the third player gets the first job. There are three possible cases. In the first, the first player gets the second job (Case 4.1), in the second case, the second player gets the second job (Case 4.2) and in the third case the third player gets the second job (Case 4.3).

\subsubsection*{Case 4.1}
In this case, $M$ has an approximation ratio of $1+x$.
\[
\begin{pmatrix}
\infty &x^{\textcolor{blue}{*}}              &\infty               &{\varepsilon^{---}}^{\textcolor{blue}{*}}\\
1      &{\varepsilon^{-}}  &\varepsilon^{-}      &\infty\\
{1-\varepsilon'}^{\textcolor{blue}{*}}       &\varepsilon^{--} &\varepsilon^{--}     &\infty\\
\end{pmatrix}
\underbrace{\xrightarrow{}}_{\text{By Lemma \ref{lemma3}}}
\quad
\begin{pmatrix}
\infty &x^{\textcolor{blue}{*}}              &\infty               &{1}^{\textcolor{blue}{*}\textcolor{red}{@}}\\
1^{\textcolor{red}{@}}      &{\varepsilon^{-}}  &\varepsilon^{-}      &\infty\\
{1-\varepsilon'}^{\textcolor{blue}{*}}       &{\varepsilon^{--}}^{\textcolor{red}{@}} &{\varepsilon^{--}}^{\textcolor{red}{@}}     &\infty\\
\end{pmatrix}
\]

\subsubsection*{Case 4.2}
In this case, $M$ has an approximation ratio which can be made arbitrarily large.
\[
\begin{pmatrix}
\infty &x             &\infty               &{\varepsilon^{---}}^{\textcolor{blue}{*}}\\
1      &{\varepsilon^{-}}^{\textcolor{blue}{*}}  &\varepsilon^{-}      &\infty\\
{1-\varepsilon'}^{\textcolor{blue}{*}}       &\varepsilon^{--} &\varepsilon^{--}     &\infty\\
\end{pmatrix}
\underbrace{\xrightarrow{}}_{\text{By Lemma \ref{lemma1}}}
\quad
\begin{pmatrix}
\infty &x             &\infty               &{\varepsilon^{---}}^{\textcolor{blue}{*}\textcolor{red}{@}}\\
1      &{\varepsilon^{-}}^{\textcolor{blue}{*}}  &\varepsilon^{-}      &\infty\\
{0}^{\textcolor{blue}{*}\textcolor{red}{@}}       &{\varepsilon^{--}}^{\textcolor{red}{@}} &{\varepsilon^{--}}^{\textcolor{red}{@}}     &\infty\\
\end{pmatrix}
\]

This results in an approximation ratio of $\frac{\varepsilon^-}{2\cdot \varepsilon^{--}}$ or, $\frac{x}{2\cdot \varepsilon^{--}}$.
The second approximation is achieved when the first player gets the second job and not the second player (after the above transition).

\subsubsection*{Case 4.3}

In this case there are two possible cases, the first is that the third job is allocated to the second player (Case 4.3.1) and the second is that the third job is allocated to the third player(Case 4.3.2). 

\subsubsection*{Case 4.3.1}
\[
\begin{pmatrix}
\infty &x             &\infty               &{\varepsilon^{---}}^{\textcolor{blue}{*}}\\
1      &{\varepsilon^{-}} &{\varepsilon^{-}}^{\textcolor{blue}{*}}      &\infty\\
{1-\varepsilon'}^{\textcolor{blue}{*}}       &{\varepsilon^{--}}^{\textcolor{blue}{*}} &\varepsilon^{--}     &\infty\\
\end{pmatrix}
\underbrace{\xrightarrow{}}_{\text{By Lemma \ref{lemma1}}}
\quad
\begin{pmatrix}
\infty &x             &\infty               &{\varepsilon^{---}}^{\textcolor{blue}{*}\textcolor{red}{@}}\\
1      &{\varepsilon^{-}} &{\varepsilon^{-}}^{\textcolor{blue}{*}}      &\infty\\
{0}^{\textcolor{blue}{*}\textcolor{red}{@}}       &{0}^{\textcolor{blue}{*}\textcolor{red}{@}} &{\varepsilon^{--}}^{\textcolor{red}{@}}     &\infty\\
\end{pmatrix}
\]

In this case, $M$ has an approximation ratio of $\frac{\varepsilon^{-}}{\varepsilon^{--}}$ which can be made arbitrarily large.

\subsubsection*{Case 4.3.2}

\[
\begin{pmatrix}
\infty &x             &\infty               &{\varepsilon^{---}}^{\textcolor{blue}{*}}\\
1      &{\varepsilon^{-}} &{\varepsilon^{-}}      &\infty\\
{1-\varepsilon'}^{\textcolor{blue}{*}}       &{\varepsilon^{--}}^{\textcolor{blue}{*}} &{\varepsilon^{--}}^{\textcolor{blue}{*}}     &\infty\\
\end{pmatrix}
\underbrace{\xrightarrow{}}_{\text{By Lemma \ref{lemma1}}}
\quad
\underbrace{\begin{pmatrix}
\infty &x             &\infty               &{\varepsilon^{---}}^{\textcolor{blue}{*}}\\
1      &\infty &\infty     &\infty\\
{1-\varepsilon'}     &{\varepsilon^{--}} &{\varepsilon^{--}}    &\infty\\
\end{pmatrix}}_{=F_4}
\]

By Lemma \ref{lemma1}, the second player does not get the first job. Thus, since $M$ has a finite approximation ratio, the third player gets it. There are two possible cases. The first is that the first player takes the second job (Case 4.3.2.1) and the second case, that the third player does (Case 4.3.2.2).

\subsubsection*{Case 4.3.2.1}
In this case, $M$ has an approximation ratio of $1+x$.
\[
\begin{pmatrix}
\infty &x^{\textcolor{blue}{*}}             &\infty               &{\varepsilon^{---}}^{\textcolor{blue}{*}}\\
1      &\infty &\infty     &\infty\\
{1-\varepsilon'}     &{\varepsilon^{--}} &{\varepsilon^{--}}    &\infty\\
\end{pmatrix}
\underbrace{\xrightarrow{}}_{\text{By Lemma \ref{lemma3}}}
\quad
\begin{pmatrix}
\infty &x^{\textcolor{blue}{*}}             &\infty               &{1}^{\textcolor{blue}{*}\textcolor{red}{@}}\\
1      &\infty &\infty     &\infty\\
{1-\varepsilon'}^{\textcolor{red}{@}}     &{\varepsilon^{--}}^{\textcolor{red}{@}} &{\varepsilon^{--}}^{\textcolor{red}{@}}    &\infty\\
\end{pmatrix}
\]

\subsubsection*{Case 4.3.2.2}
\[
\begin{pmatrix}
\infty &x             &\infty               &{\varepsilon^{---}}^{\textcolor{blue}{*}}\\
1      &\infty &\infty     &\infty\\
{1-\varepsilon'}^{\textcolor{blue}{*}}     &{\varepsilon^{--}}^{\textcolor{blue}{*}} &{\varepsilon^{--}}^{\textcolor{blue}{*}}    &\infty\\
\end{pmatrix}
\underbrace{\xrightarrow{}}_{\text{By Lemma \ref{lemma3}}}
\quad
\begin{pmatrix}
\infty &x           &\infty               &{\varepsilon^{---}}^{\textcolor{blue}{*}\textcolor{red}{@}}\\
1      &\infty &\infty     &\infty\\
{1-\varepsilon'}    &{1}&{\varepsilon^{--}}   &\infty\\
\end{pmatrix}
\]

By Lemma \ref{lemma4}, If the third player gets the second job, he also gets the first job, this case is analyzed in 4.3.2.2.1. Otherwise, the first player gets the second job, this is analyzed in 4.3.2.2.2.

\subsubsection*{Case 4.3.2.2.1}
In this case, $M$ has an approximation ratio of $\frac{2+x}{x}$. 

\[
\begin{pmatrix}
\infty &x           &\infty               &{\varepsilon^{---}}^{\textcolor{blue}{*}}\\
1      &\infty &\infty     &\infty\\
{1-\varepsilon'}^{\textcolor{blue}{*}}    &{1}^{\textcolor{blue}{*}}&{\varepsilon^{--}}^{\textcolor{blue}{*}}   &\infty\\
\end{pmatrix}
\underbrace{\xrightarrow{}}_{\text{By Lemma \ref{lemma3}}}
\quad
\begin{pmatrix}
\infty &x^{\textcolor{red}{@}}           &\infty               &{\varepsilon^{---}}^{\textcolor{blue}{*}\textcolor{red}{@}}\\
1^{\textcolor{red}{@}}      &\infty &\infty     &\infty\\
{1-\varepsilon'}^{\textcolor{blue}{*}}    &{1}^{\textcolor{blue}{*}}&{x}^{\textcolor{blue}{*}\textcolor{red}{@}}   &\infty\\
\end{pmatrix}
\]

\subsubsection*{Case 4.3.2.2.2}
In this case, $M$ has an approximation ratio of $1+x$. 
\[
\begin{pmatrix}
\infty &x^{\textcolor{blue}{*}}           &\infty               &{\varepsilon^{---}}^{\textcolor{blue}{*}}\\
1      &\infty &\infty     &\infty\\
{1-\varepsilon'}   &{1}&{\varepsilon^{--}}   &\infty\\
\end{pmatrix}
\underbrace{\xrightarrow{}}_{\text{By Lemma \ref{lemma3}}}
\quad
\begin{pmatrix}
\infty &x^{\textcolor{blue}{*}}           &\infty               &{1}^{\textcolor{blue}{*}\textcolor{red}{@}}\\
1^{\textcolor{red}{@}}      &\infty &\infty     &\infty\\
{1-\varepsilon'}   &{1}^{\textcolor{red}{@}}&{\varepsilon^{--}}^{\textcolor{red}{@}}   &\infty\\
\end{pmatrix}
\]
\subsubsection*{Case 5}
In this case, $M$ has an approximation ratio of $\frac{\varepsilon^{+}}{\varepsilon^{-}}$ which can be made arbitrarily large.
\[
\begin{pmatrix}
\infty &x               &\infty               &{\varepsilon^{---}}^{\textcolor{blue}{*}}\\
1      &\varepsilon^{-} &\varepsilon^{-}      &\infty\\
1^{\textcolor{blue}{*}}       &{\varepsilon^{+}}^{\textcolor{blue}{*}}  &\varepsilon^{--}     &\infty\\
\end{pmatrix}
\underbrace{\xrightarrow{}}_{\text{By Lemma \ref{lemma1}}}
\quad
\begin{pmatrix}
\infty &x               &\infty               &{\varepsilon^{---}}^{\textcolor{blue}{*}\textcolor{red}{@}}\\
1      &{\varepsilon^{-}}^{\textcolor{red}{@}} &\varepsilon^{-}      &\infty\\
0^{\textcolor{blue}{*}\textcolor{red}{@}}       &{\varepsilon^{+}}^{\textcolor{blue}{*}}  &{\varepsilon^{--}}^{\textcolor{red}{@}}     &\infty\\
\end{pmatrix}
\]


\bibliography{ref}
\bibliographystyle{plain}




\end{document}